\documentclass[
reprint,
superscriptaddress,
 amsmath,amssymb,
 aps,
pra,
floatfix,
]{revtex4-1}

\usepackage{graphicx}
\usepackage{dcolumn}
\usepackage{bm}
\usepackage{float}
\usepackage{braket}
\usepackage{xcolor}
\usepackage{breqn}

\usepackage{xcolor}

\usepackage{mathtools}
\usepackage{amsmath}
\newcommand\numberthis{\addtocounter{equation}{1}\tag{\theequation}}

\begin{document}
\hfuzz=2pt

\title{Symmetric Bloch oscillations of matter waves}

\author{Zachary Pagel}
\email{zpagel@berkeley.edu}
\affiliation{Department of Physics$,$ University of California Berkeley$,$ Berkeley$,$ CA 94720$,$ USA}
\author{Weicheng Zhong}
\affiliation{Department of Physics$,$ University of California Berkeley$,$ Berkeley$,$ CA 94720$,$ USA}
\author{Richard H. Parker}
\affiliation{Department of Physics$,$ University of California Berkeley$,$ Berkeley$,$ CA 94720$,$ USA}
\author{Christopher T. Olund}
\affiliation{Department of Physics$,$ University of California Berkeley$,$ Berkeley$,$ CA 94720$,$ USA}
\author{Norman Y. Yao}
\affiliation{Department of Physics$,$ University of California Berkeley$,$ Berkeley$,$ CA 94720$,$ USA}
\affiliation{Materials Science Division$,$ Lawrence Berkeley National Laboratory$,$ Berkeley$,$ CA 94720$,$ USA}
\author{Holger M\"{u}ller}
\email{hm@berkeley.edu}
\affiliation{Department of Physics$,$ University of California Berkeley$,$ Berkeley$,$ CA 94720$,$ USA}

\date{\today}

\begin{abstract}
Cold atoms in an optical lattice provide an ideal platform for studying Bloch oscillations
. Here, we extend Bloch oscillations to two superposed optical lattices that are accelerated away from one another, and for the first time show that these symmetric Bloch oscillations can split, reflect and recombine matter waves coherently. 
Using the momentum parity-symmetry of the Hamiltonian, we map out the energy band structure of the process and show that superpositions of momentum states are created by adiabatically following the ground state of the Hamiltonian. The relative phase and velocity of the two lattices completely determines the trajectories of different branches of the matter wave. Experimentally, we demonstrate symmetric Bloch oscillations using cold Cesium atoms where 
we form interferometers with up to $240\hbar k$ momentum splitting, one of the largest coherent momentum splittings achieved to date. This work has applications in macroscopic tests of quantum mechanics, measurements of fundamental constants, and searches for new physics. 
\end{abstract}
\maketitle


\section{\label{sec:level1}Introduction}
Bloch oscillations and the Wannier--Stark ladder of matter waves in a periodic potential were first studied in the context of electrons in crystals in the presence of a homogenous electric field \cite{Bloch original, Wannier original}. Their counterintuitive nature---that a constant electric field should lead to an ac current---triggered a debate about their existence \cite{e oscillate, comment e} and led to the formulation of criteria for their observability \cite{e dynamics EM}.   
Bloch oscillations were first experimentally observed in semiconductor superlattices \cite{Bloch observe 1, Bloch observe 2}, and have since been studied in a wide variety of physical systems ranging from Bloch oscillations of light \cite{Bloch light, Bloch light 2} to cold atoms \cite{Salomon, Raizen}. 
Bloch oscillations are particularly useful in matter wave interferometers, which have found widespread applications in precision measurements of fundamental constants \cite{alpha, LKB alpha, tino G, kasevich G}, tests of the weak equivalence principle \cite{wuhan WEP, Tino WEP} and dark energy theories \cite{chameleon 1, chameleon 2}, as well as precision gravimetry \cite{wuhan gravity, minig} and gradiometry \cite{kasevich gradiometry}. 

Matter wave interferometers use optical lattices to coherently transfer momentum, allowing one to split a matter wave between different spatial trajectories, then later recombine them and create interference. 
The measured phase can be increased by using larger momentum splitting between the trajectories \cite{alpha, LKB alpha}; Bloch oscillations enable such a process \cite{bloch atoms salomon, biraben Bloch} and have recently shown to coherently transfer the momentum of more than $10^4$ photons to the atoms \cite{20s}. With two superposed lattices that are independently accelerated, it might even be possible to realize large-momentum-transfer beam splitters for matter waves, by performing Bloch oscillations of two different velocity classes of atoms simultaneously \cite{Berman BBS}. However, this process has never been demonstrated. Near velocity degeneracy of the two accelerated lattices, it was expected that non-adiabatic effects would prevent coherent ground state dynamics. Instead, Bloch oscillations have only been used to accelerate atoms after an initial momentum splitting was already made with Bragg diffraction \cite{BBB,alpha}, resulting in up to $408\hbar k$ momentum splittings \cite{Abend Thesis,New Abend LMT}, where $k$ is the wavevector of the laser. 

Here, we show that Bloch oscillations of atoms in two symmetrically accelerated lattices can remain adiabatic and coherent even as the two lattices pass through velocity degeneracy. Theoretically, we show that it is possible to split, reflect, and recombine atoms simply by allowing them to adiabatically follow the ground state of the Hamiltonian while accelerating the two lattices. The dynamics result in symmetric Bloch oscillations where the matter wave is in a coherent superposition of interacting with each of the two lattices, and the relative phase and velocity of the two lattices completely determines the trajectories of different branches of the matter wave. Experimentally, we demonstrate symmetric Bloch oscillations and realize $240\,\hbar k$ coherent momentum splitting of a superposition state as well as interferometry with nearly fully-guided matter waves. 

Using only accelerated lattices for momentum transfer is desirable for a number of reasons. In comparison with resonant processes such as Bragg diffraction, 1) the dynamics are adiabatic, and can therefore be much more efficient per $\hbar k$ momentum transfer, 2) the processes require less laser power, 3) the velocity class of atoms addressed can be larger, relaxing temperature requirements on atom clouds, and 4) the optical lattices prevent thermal expansion of the atom cloud, further relaxing temperature requirements. As a result, symmetric Bloch oscillations can find applications in next-generation precision measurements of fundamental constants, searches for gravitational waves, and searches for new physics \cite{alpha ann, Tino WEP, zhan GW, graham resonant, chameleon 2}.

\begin{figure}[t]
\centering
\includegraphics[width=3in]{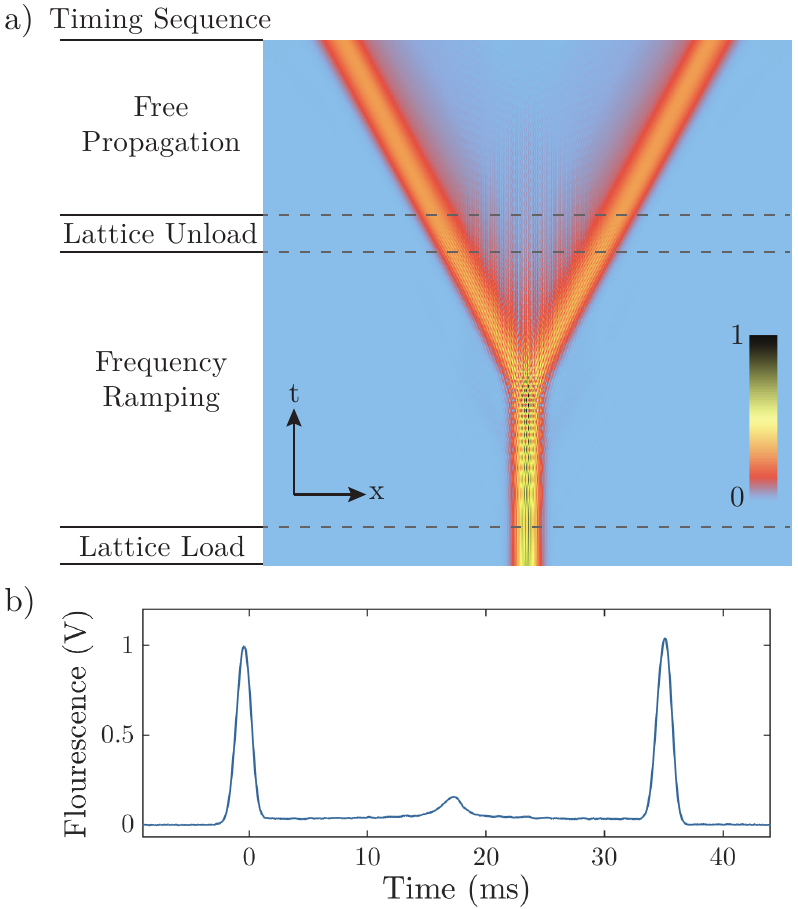}
\caption{a)  Density plot of $|\psi(x,t)|$ from numerical simulation of a symmetric beamsplitter with a lattice depth $U_0 = 1.3\,E_r$ and ramp rate $r = 0.2\,\omega_r^2$. Timing sequence is indicated on the left. The initial wavefunction is a Heisenberg-limited gaussian wavepacket with velocity spread 0.05$\,v_r$, corresponding to our experimental temperature. Frequencies are ramped for one Bloch period, corresponding to a $4\hbar k$ splitting between arms. b) Experimental time of flight fluorescence trace showing an efficient $60\hbar k$ beamsplitter with a ramp rate $r = 0.26\:\omega_r^2$, and a lattice depth of around $1.5\:E_r$.}
\label{fig:fig1}
\end{figure}

Section \ref{sec:theory} presents a theoretical treatment of the Hamiltonian and the resulting dynamics. The Hamiltonian is symmetric under momentum inversion, allowing one to simultaneously diagonalize the Hamiltonian in momentum parity and energy. For the beamsplitter process described above, we show that an atom adiabatically follows the even-parity ground state of the Hamiltonian. The momentum-parity basis is then used to study effects such as non-adiabatic losses, dynamics while ramping the lattices through velocity degeneracy, and effects from different experimental imperfections. 

In section \ref{sec:experiment}, we describe how we implement symmetric Bloch oscillations experimentally. We use the relative phase between the two lattices to control the populations in the two lattices after ramping through velocity degeneracy; in effect, this creates a fully tunable matter-wave switch each time the lattices cross through velocity degeneracy. We demonstrate the first interferometers created only using accelerated lattices, including a Mach--Zehnder (MZ) interferometer with a momentum splitting of up to $240\hbar k$. Prior to this work, the largest momentum transfer from a single beamsplitter operation was $24\hbar k$ \cite{holger Bragg}. In order to confirm that symmetric Bloch oscillations are first-order phase coherent, we implement a differential measurement between two simultaneous MZ interferometers and see a stable phase between the interferometer outputs.


\section{\label{sec:theory}Theory}

When two superposed optical lattices are far apart in velocity, it is well known that atoms can undergo efficient Bloch oscillations in either of the lattices \cite{BBB,alpha,Abend Thesis,New Abend LMT}. Near velocity degeneracy, however, it was previously expected that near-resonant effects from the second lattice would cause too large of a perturbation to the standard Bloch oscillation dynamics to permit an efficient beamsplitter. We first derive a unitary transformation that isolates the relevant dynamics (Sec.~\ref{sec:levelA}), and then show that the effects of the perturbation terms can remain small within the rotating wave approximation under certain conditions (Sec.~\ref{sec:RWA ramp}). Throughout the analysis, it is useful to stress the parallels between Bloch oscillations in a single lattice (SLBO) and Bloch oscillations in two lattices which we call dual-lattice Bloch oscillations (DLBO). The simplified DLBO Hamiltonian is nearly identical to the SLBO Hamiltonian, differing only in being invariant under momentum inversion. As a result, the eigenstates of DLBO are symmetric and anti-symmetric in momentum space.

We then study non-adiabatic loss mechanisms, which include standard Landau--Zener tunneling due to avoided level crossings as well as higher-order transitions which are possible due to perturbation terms dropped in the rotating wave approximation (Sec.~\ref{sec:LZ ramp}). These conditions are combined to place limits on the permissible lattice accelerations and lattice depths, and in total they allow for the DLBO to approach 100\% efficiency in the limit of slowly accelerated lattices (Sec.~\ref{sec:Combined ramp}). The dynamics are also discussed for lattices that are ramped through velocity degeneracy, showing that an offset laser phase can be used to coherently control the output population in the two lattices (Sec.~\ref{sec:degeneracy}). Lastly, we discuss some important experimental requirements in order to realize these methods in the laboratory (Sec.~\ref{eh}), and supporting material is left for the Appendices (Sec.~\ref{sec:appendix}).

\subsection{\label{sec:levelA}Hamiltonian and unitary transformation}

SLBO are most easily studied using a coordinate system that is comoving with the accelerating lattice \cite{Clade, Kolovsky, Unitary}, and a unitary transformation can be used to boost the Hamiltonian between the atom's inertial frame and the accelerating lattice frame \cite{Kovachy guided interferometer, Unitary}. For DLBO, it is not possible to transform to a coordinate system that is simultaneously comoving with both lattices. Instead, using a basis of momentum states it is possible to independently transform each momentum state so that positive (negative) momentum states are boosted to a coordinate system comoving with the positively (negatively) accelerating lattice. This unitary transformation is shown to capture the core coherent dynamics of DLBO. The analysis that follows is relevant for zero temperature atoms comoving with the initially degenerate lattices: a similar analysis can be explored for atoms with a small initial velocity, and the band structure of the Hamiltonian can still be studied. One finds that any initial velocity breaks the parity symmetry discussed in the following sections and leads to asymmetric dynamics. A full analysis is beyond the scope of this paper. 

\begin{figure}[t!]
\centering
\includegraphics[width=3.3in]{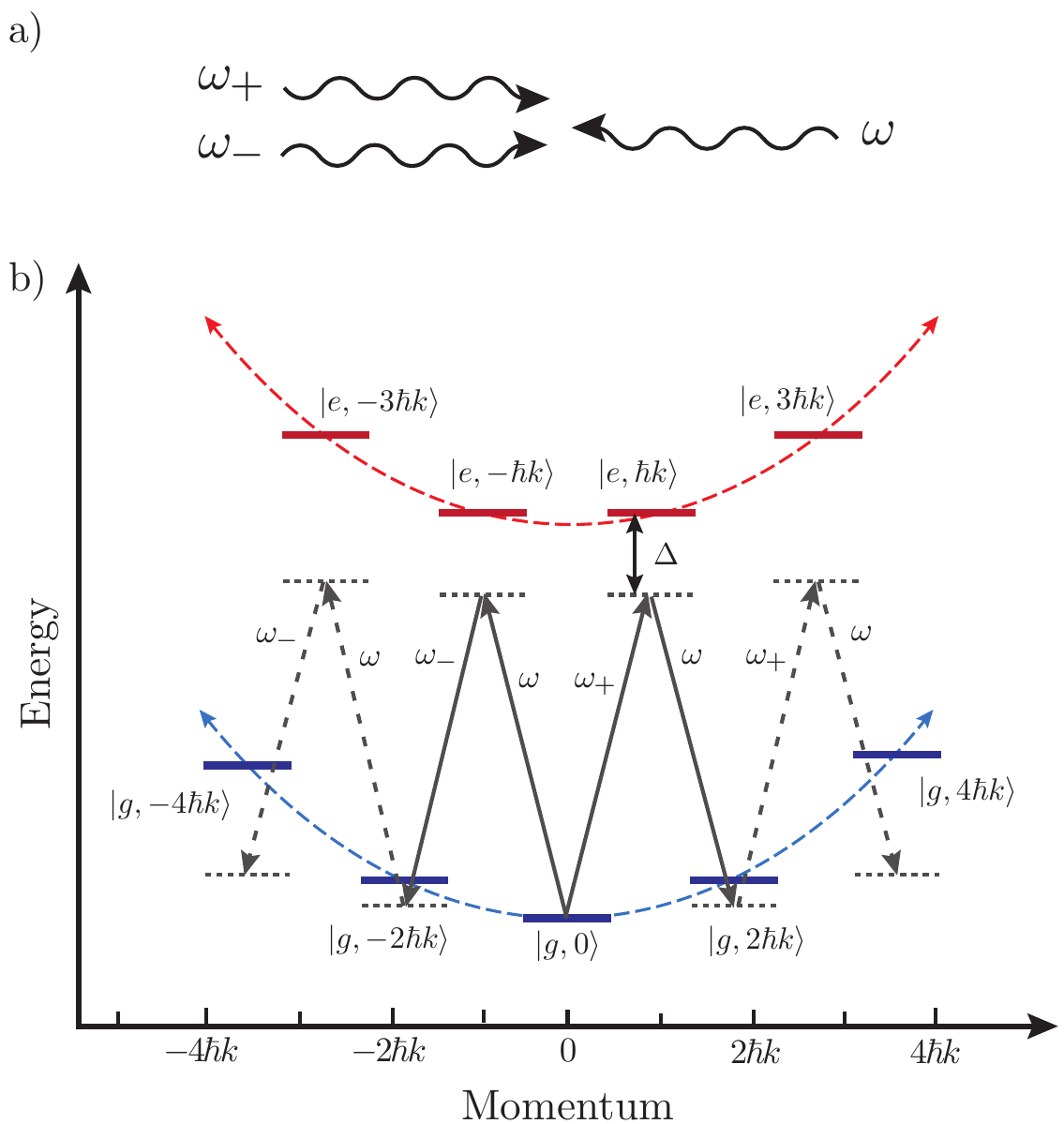}
\caption{a) Counter-propagating lasers form two superposed optical lattices. The frequency differences are $\omega_+-\omega = \omega - \omega_- = \omega_m$. b) Energy-momentum level diagram showing relevant atomic states. The lasers drive two-photon transitions between neighboring momentum states such that the atom remains in the same internal ground state. The detuning from the excited states $\Delta$ (many GHz) is much larger than the separation between adjacent ground states (few kHz). As the modulation frequency $\omega_m$ is swept away from zero, the lasers sweep past a succession of two-photon transitions between adjacent ground states. Off-resonant transitions driven by the extra oscillating terms in the Hamiltonian (Eq.~\ref{new Ham}) are omitted for clarity. }
\label{fig:levels}
\end{figure}

We begin with a Hamiltonian containing the AC Stark shift of two superposed optical lattices that are far detuned from single-photon transitions (see Fig.~\ref{fig:levels}). Experimentally, the lattices are realized with one upward-propagating laser frequency $\omega_1$, and two downward-propagating frequencies $\omega_2 \pm \omega_m(t)$. We work in the frame of reference where $\omega_1 = \omega_2 = \omega$, and denote $\omega_{\pm} = \omega \pm \omega_m(t)$. 
The relative speed of the two lattices is given by $\omega_m(t)/k$, where $k$ is the wave number of the laser defined as $k = \omega/c$. Two-photon transitions leave atoms in the same internal state but different external momentum states. After adiabatic elimination of the excited state, the Hamiltonian for an atom in these two optical lattices can be written as:

    \begin{multline}
        \label{BBS Hamiltonian}
        H_{\rm BBS}(t) = \frac{\hat{p}^{2}}{2m} + \frac{U_0}{2}\bigg(\cos\left[2k_+\hat{x} + \int_0^t\omega_m\left(t'\right)dt' + \phi_{1}\right] \\ + \cos\left[2k_-\hat{x} - \int_0^t\omega_m\left(t'\right)dt' + \phi_{2}\right]\bigg) \\
        =\frac{\hat{p}^{2}}{2m} + U_0\, \cos\left[2k\hat{x}\right]\cos\left[\int_0^t\omega_m\left(t'\right)dt' + \phi_{0}\right].
    \end{multline}
Constant terms are dropped in the second form, which will be used for analytics and simulation. The wave numbers $k_+ = \omega_+/c$ and $k_- = \omega_-/c$ are nearly identical to $k$, so we approximate $k_+\approx k_-\approx k$ in the second form as well. For Cs atoms separated by $n=1000$ photon momenta, $k_+$, $k_-$, and $k$ differ by less than one part in $10^{8}$. The phases $\phi_0$, $\phi_1$, and $\phi_2$ are offsets between counter-propagating lasers at time $t=0$. The lattice depth $U_0 = \hbar \Omega_R^2/(2\Delta)$ is the AC Stark shift for a single, far-detuned lattice 
\cite{Clade}, where $\Delta$ is the detuning from the excited state and $\Omega_R$ is the on-resonance Rabi frequency between the ground and excited states. The integral $\int_0^t\omega_m\left(t'\right)dt'$ keeps track of the phase evolution of the lattice for time dependent frequencies. Specializing to linear frequency ramps with rate $r$, the modulation frequency can be written as $\omega_m(t) = r t$ so that the lattices are velocity degenerate at time $t=0$ and $\int_0^t\omega_m\left(t'\right)dt' = r t^2/2$. This ramp rate corresponds to an acceleration $a = r/2 k$.

We now write the Hamiltonian in a momentum-state basis 
$\ket{l}$, where $l$ is an integer that labels the basis states such that the state $\ket{l}$ has $2 l \hbar k$ momentum. 
Plane-wave basis states are a good approximation to initial atomic states when the velocity spread is much smaller than the recoil velocity $v_r=\hbar k/m$. 
Projected into this basis, the Hamiltonian is:

\begin{multline}
    \label{BBS Hamiltonian kets}
    H = \sum_{l=-\infty}^{\infty}\Bigg(\frac{(2 l \hbar k)^2}{2 m}\ket{l}\!\!\bra{l} \\ + U_0 \cos\left(\frac{r t^2}{2} + \phi_0\right)\left(\ket{l}\!\!\bra{l+1} + \ket{l}\!\!\bra{l-1}\right)\Bigg)
\end{multline}

The unitary transformation used to boost the different momentum states in this Hamiltonian is given by:
\begin{align*}
    \label{unitary}
    U &= \sum_{l=-\infty}^{\infty} e^{i \frac{d(t) |\hat{p}|}{\hbar}}e^{i \frac{\theta(t)}{\hbar}}\ket{l}\!\!\bra{l} \numberthis
\end{align*}
where $d(t) \equiv a t^2/2 + \phi_0/k$ and $\theta(t) \equiv m a^2 t^3/6$. The first term corresponds to the position translation operator, and the absolute value sign ensures that positive momentum states are translated with the positive-moving lattice while negative momentum states are translated with the negative-moving lattice. The $d(t)$ term in Eq.~(\ref{unitary}) also absorbs the offset phase $\phi_0$ into the definition of the basis states. The $\theta(t)$ in Eq.~(\ref{unitary}) corresponds to a global energy shift to each state such that the energy of the ground states comoving with either of the lattices stays near zero at all times \cite{Kovachy guided interferometer}. 
See Appendix \ref{sec:SLBO Ham} for the analogous treatment of the SLBO Hamiltonian.




The transformed Hamiltonian $H' = U H U^{\dagger} + i\hbar\frac{dU}{dt}U^{\dagger}$ is:
\begin{multline}
    \label{new Ham}
    H' = \sum_{l\neq 0}\Bigg[\frac{(2 |l| \hbar k - F t)^2}{2 m}\ket{l}\!\!\bra{l} \\ + \frac{U_0}{2}\left(1 + e^{i s_l(r t^2 + 2 \phi_0)}\right)\ket{l}\!\!\bra{l+1}
    \\ + \frac{U_0}{2}\left(1 + e^{-i s_l(r t^2 + 2 \phi_0)}\right) \ket{l}\!\!\bra{l-1}\Bigg]
    \\ + \frac{(F t)^2}{2 m}\ket{0}\!\!\bra{0} 
     + \frac{U_0}{2}\left(1 + e^{- i (r t^2 + 2 \phi_0)}\right)\left(\ket{0}\!\!\bra{1}
    + \ket{0}\!\!\bra{-1} \right)
\end{multline}
where $s_l \equiv l/|l|$ is the sign of the momentum state, and the force $F = r m/2 k$ is adapted from the standard treatment of SLBO \cite{Clade}. 




The nearest-neighbor coupling terms proportional to $|l\rangle\!\langle l\pm 1|$ include both a stationary term and an oscillating term. In a two-level system, oscillating coupling terms of this type can be dropped under a rotating wave approximation (RWA) provided the terms time-average to zero on the relevant timescale of the dynamics. Here, the couplings between neighboring momentum states can be treated with an analogous RWA to arrive at the reduced DLBO Hamiltonian:
\begin{multline}
    \label{final Ham}
    H_{\text{DLBO}} = \sum_{l=-\infty}^{l=\infty}\frac{(2 |l| \hbar k - F t)^2}{2 m}\ket{l}\!\!\bra{l} \\ + \frac{U_0}{2}\left(\ket{l}\!\!\bra{l+1} + \ket{l}\!\!\bra{l-1}\right)
\end{multline}
The validity of this RWA is discussed in Sect.~\ref{sec:RWA ramp}, where we derive bounds on the ramp rate for which the Hamiltonian in Eq.~(\ref{final Ham}) is valid.


\begin{figure}[t]
\centering
\includegraphics[width=3.15in]{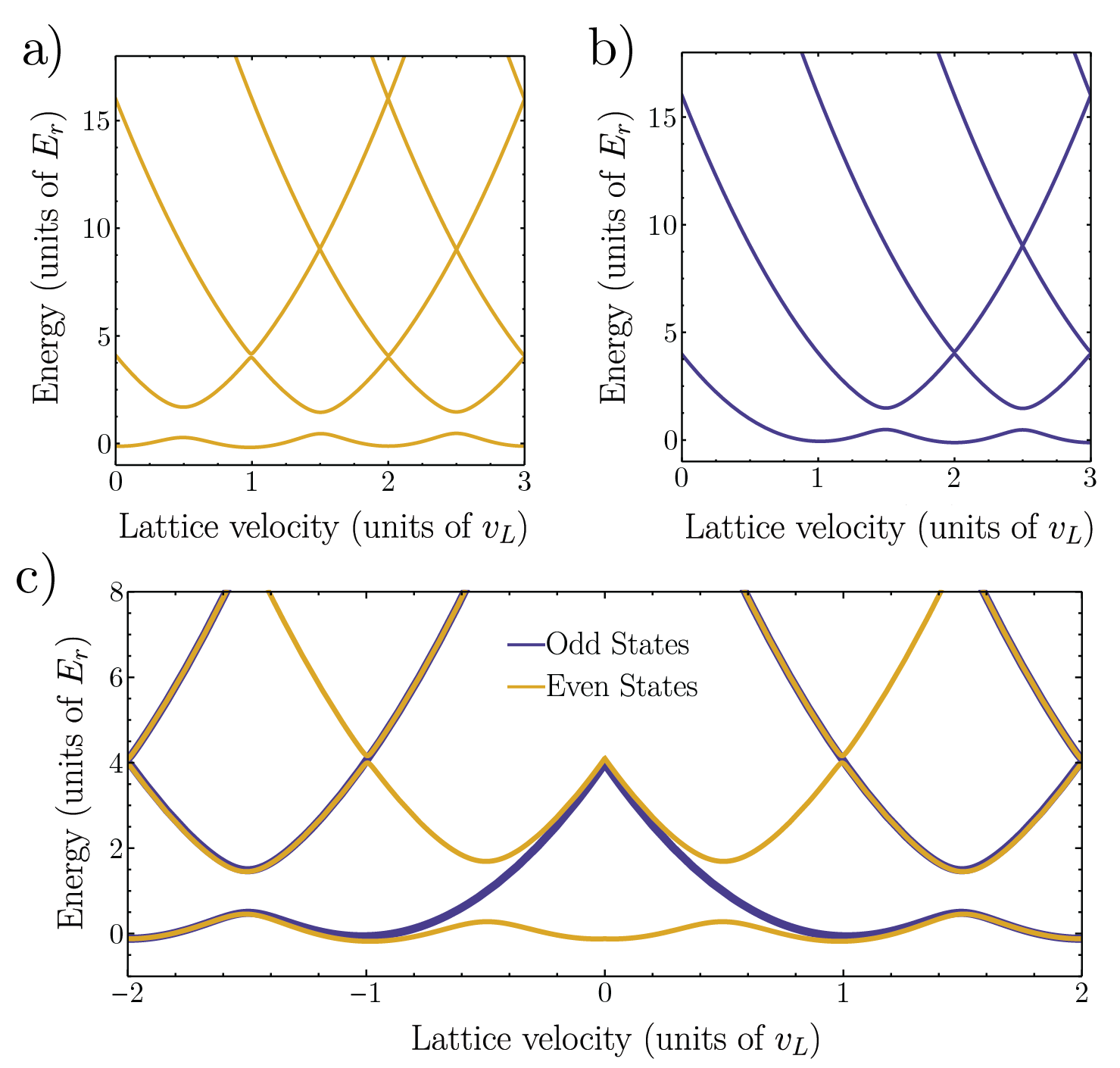}
\caption{Energy band structures of the reduced Hamiltonian (\ref{final Ham}) as a function of the lattice velocity, using a lattice depth $U_0 = 1 E_r$. The lattice velocity is defined as $v_L = r t$, such that the two lattices at time $t$ have velocities $\pm v_L$. a) Even-parity and b) Odd-parity energy eigenvalues starting from velocity degeneracy. c) Combined band structure as lattices are ramped through velocity degeneracy.}
\label{fig:bands}
\end{figure}

The DLBO Hamiltonian in Eq.~(\ref{final Ham}) and the SLBO Hamiltonian derived in Appendix \ref{sec:SLBO Ham} are nearly identical; the only difference is the absolute value $|l|$ in the kinetic energy term for $H_{\text{DLBO}}$, which makes $H_{\text{DLBO}}$ symmetric under momentum inversion. This symmetry is already apparent in the original Hamiltonian (\ref{BBS Hamiltonian}), which commutes with a momentum inversion operator. 
Using a basis of momentum eigenstates that are also eigenstates of momentum-parity, the even- and odd-parity states are decoupled.

Figure \ref{fig:bands} (a,b) shows the energy band structure over time of the Hamiltonian (\ref{final Ham}) for even- and odd-parity states, respectively, where the two lattices are ramped away from velocity degeneracy beginning at time $t=0$. 
The energy bands are calculated by finding eigenvalues of a truncated version of the Hamiltonian in Eq.~(\ref{final Ham}) as a function of time. Note that in plotting the energy bands in Fig.~\ref{fig:bands}c, for negative times we use the substitution $d(t) \rightarrow -d(t)$ in Eq.~(\ref{unitary}) in order to use the coordinate frame comoving with the lattices driving amplitude towards zero momentum instead of driving amplitude away from zero momentum.

A beamsplitter can be understood as an atom adiabatically following the even-parity ground state of the Hamiltonian (\ref{final Ham}), and higher efficiency beamsplitters can be achieved by making the process more adiabatic. At every time $t=(a+1/2)T_B$ for integers $a\geq0$, where the Bloch period $T_B=8\omega_r/r$ and the recoil frequency $\omega_r = \hbar k^2/(2m)$,  there is a level crossing such that the even-parity state receives an additional $4\hbar k$ momentum splitting; the positive momentum component of the even state acquires an additional $+2\hbar k$ momentum and the negative momentum component acquires an additional $-2\hbar k$ momentum. This is the momentum-symmetric analogue of SLBO in the ground Bloch band, where atoms receive $2\hbar k$ momentum at the edge of the first Brillouin zone at each avoided level crossing between the ground band and first excited band.

\subsection{\label{sec:RWA ramp}Limits on ramp rate from the rotating wave approximation}

A RWA can be used to drop the oscillating coupling terms in Eq.~(\ref{new Ham}) provided that the time-average of the oscillating term $e^{i r t^2}$ is $\ll 1$ on the relevant timescale of the dynamics, namely the duration of first level crossing between the ground even band and the first excited even band. This crossing occurs at time $t = T_B/2$, and the time interval during which the level crossing happens is given by $\Delta t = 2\sqrt{2}U_0/(\hbar r)$. A simplified form of the resulting inequality gives an upper limit on the ramp rate for which the RWA is valid:

\begin{equation}
	\label{reduced RWA condition}
r\ll 4U_0(2\sqrt{2}E_r-U_0)/\hbar^2
\end{equation}

where we define the recoil energy $E_r = \hbar \omega_r$. The RWA is therefore valid in the limit as $r\rightarrow 0$. See Appendix \ref{sec:RWA} for a full derivation of this condition.

The validity of the RWA can be further studied with numerical simulation. By solving for the evolution of $\ket{\psi(t)}$ from the Hamiltonian in Eq.~(\ref{new Ham}), the full state evolution is captured without using the RWA. We numerically integrate the Schr\"odinger equation with the Hamiltonian Eq.~(\ref{BBS Hamiltonian kets}). The initial condition is a free particle (plane-wave) momentum state which is adiabatically loaded into the lattice; the modulation frequency is then ramped to its final value, and finally the lattice is adiabatically unloaded. This state evolution can then be compared with the eigenstates of the Hamiltonian in Eq.~(\ref{final Ham}) after the RWA. Fig.~\ref{fig:LZ}a) shows the probability amplitude in the ground state of Eq.~(\ref{final Ham}) during the frequency ramping, defined as $P_0(t) = |\!\braket{+_{gs}(t)|\psi(t)}\!|^2$. The state $\ket{+_{gs}(t)}$ denotes the even-parity ground state of Hamiltonian (\ref{final Ham}) as a function of time. Fig.~\ref{fig:LZ}a) shows that the true state evolution is nearly identical to that of the ground state of the Hamiltonian in Eq.~(\ref{final Ham}), which generally holds true when Eq.~(\ref{reduced RWA condition}) is satisfied. 

To stress the parallel between SLBO and DLBO, we also plot the probability amplitude in the ground state for SLBO using eigenstates calculated from the Hamiltonian in Eq.~(\ref{Ham Bloch}) in Appendix \ref{sec:SLBO Ham}. In both SLBO and DLBO, the states pass avoided level crossings at times $t = (a+1/2)T_B$ for integer $a$, where there is mixing with the second band as well as Landau--Zener tunneling losses, which are discussed in Sec.~\ref{sec:LZ ramp}. The dual-lattice simulation doesn't project perfectly onto the ground eigenstate around time $t=0$ due to the perturbation terms dropped in the RWA.

\subsection{\label{sec:LZ ramp}Limits on ramp rate from Landau--Zener tunneling and higher-order transitions}

\begin{figure}[t]
\centering
\includegraphics[width=3.35in]{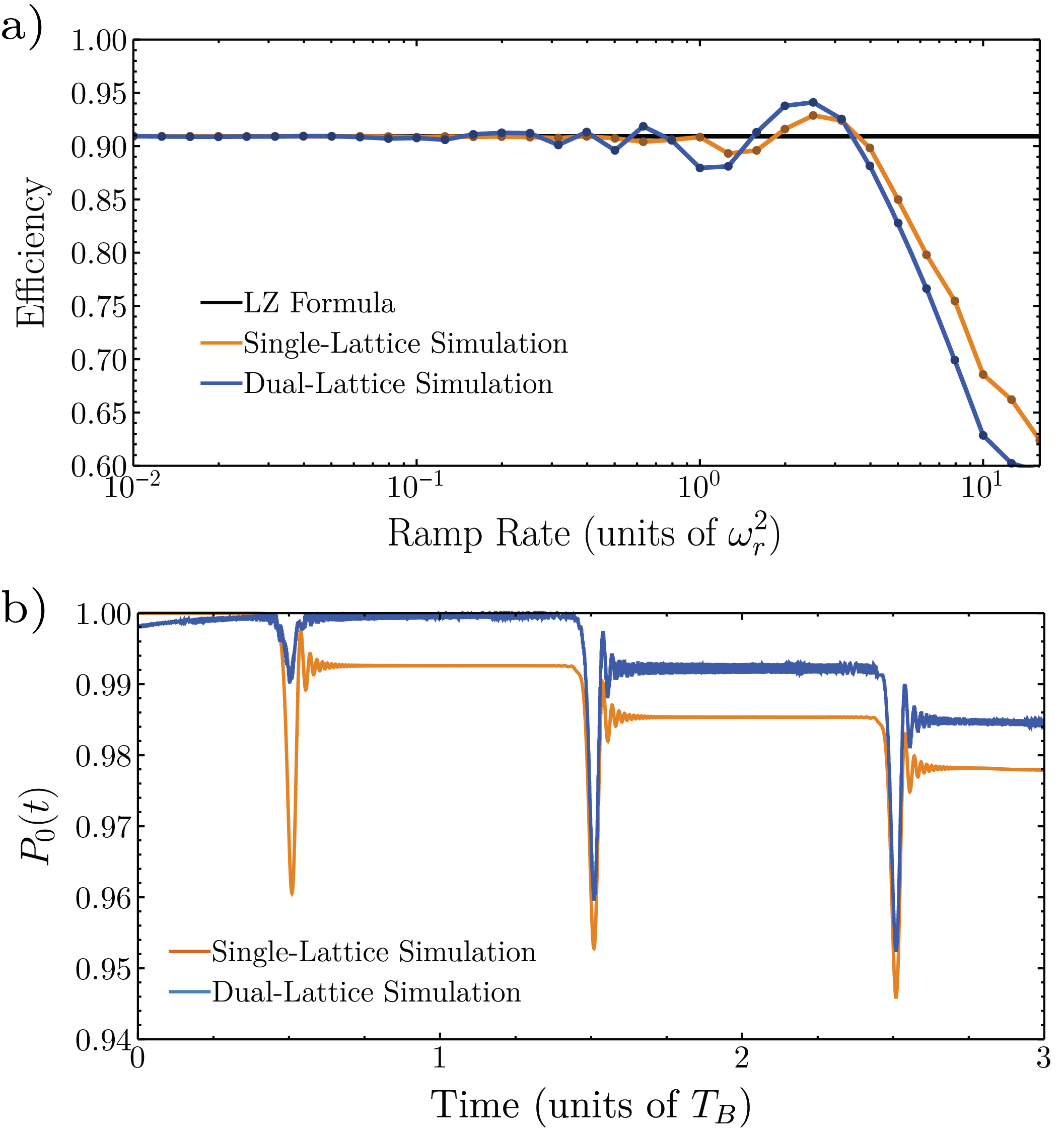}
\caption{Comparison of single-lattice and dual-lattice Bloch oscillations. a) Probability amplitude in the ground state over three Bloch periods. A lattice depth of $U_0 = 0.5 E_r$ and ramp rate $r = 0.02 \omega_r^2$ are used for both simulations. The lattice depth is intentionally chosen to be low in order to illustrate loss mechanisms for SLBO in comparison with DLBO. See text for discussion. b) Simulation of efficiencies after one level crossing. For each ramp rate, the lattice depth is chosen to keep the Landau--Zener (LZ) parameters constant at $\Gamma_{1} = \Gamma_{2} = 0.3$ such that the expected losses from the LZ formula are constant. The atom begins in the ground state at time $t=0$ with $\omega_m(t=0) = 0$ and $\phi_0 = 0$. }
\label{fig:LZ}
\end{figure}

Non-adiabatic Landau--Zener losses arise from the level crossings in Fig.~\ref{fig:bands} between the first and second even-parity energy bands. For SLBO with weak lattices and slow ramp rates, the survival probability per Bloch oscillation is given by $P_{LZ} = 1 - e^{-2\pi\Gamma_1}$ where $\Gamma_{1} = U_0^2/(4\hbar^2r)$ is the Landau--Zener parameter \cite{LZ 1, LZ 2}. For ramp rates $r < \omega_r^2$, this formula also describes losses from all level crossings of the DLBO Hamiltonian, Eq.~(\ref{final Ham}), except for the two level-crossings at $t=\pm T_B/2$. These two crossings between even-parity eigenstates have an energy gap that is increased by a factor of $\sqrt{2}$, as derived in Appendix \ref{sec:symmetrized}. The Landau--Zener parameter $\Gamma_2$ for these two crossings is therefore given by $\Gamma_2 = U_0^2/2\hbar^2r$. All subsequent crossings in DLBO have the same energy gap as SLBO and are described by the same tunneling parameter $\Gamma_{1}$. The dual-lattice beamsplitter is therefore more robust to Landau--Zener losses at the first level crossing than SLBO at a fixed lattice depth $U_0$, as shown in Fig.~\ref{fig:LZ}a).  

Fig. \ref{fig:LZ}b) shows the simulated efficiency of a single Bloch oscillation at a constant Landau--Zener parameter for both the SLBO and DLBO Hamiltonians in Eq.~(\ref{BBS Hamiltonian kets}) and (\ref{Ham Bloch lab frame}) respectively. The efficiency is defined as the total population in the desired final momentum states relative to the initial population. In order to have the same expected Landau--Zener losses for both simulations, the SLBO lattice depth is increased by a factor of $\sqrt{2}$ for each ramp rate compared to the DLBO simulation such that $\Gamma_{1} = \Gamma_{2} = 0.3$. There is asymptotic agreement with the Landau--Zener formula for ramp rates $r\ll\omega_r^2$ for both single-lattice and dual-lattice level crossings, as well as additional oscillatory behavior of the DLBO efficiency compared to the SLBO efficiency owing to the oscillatory terms dropped in the RWA.

The rotating terms being dropped in the RWA can also contribute to higher-order processes that couple amplitude from the ground band to higher energy bands, and are further discussed in Appendix \ref{sec:higher order}. The dominant loss channel is a third-order transition that couples the first and second energy levels around time $t=T_B/6$. These higher-order losses place a lower limit on the ramp rate for a fixed lattice depth, below which losses from the ground band begin to be appreciable.

\begin{figure}[t]
\centering
\includegraphics[width=3.3in]{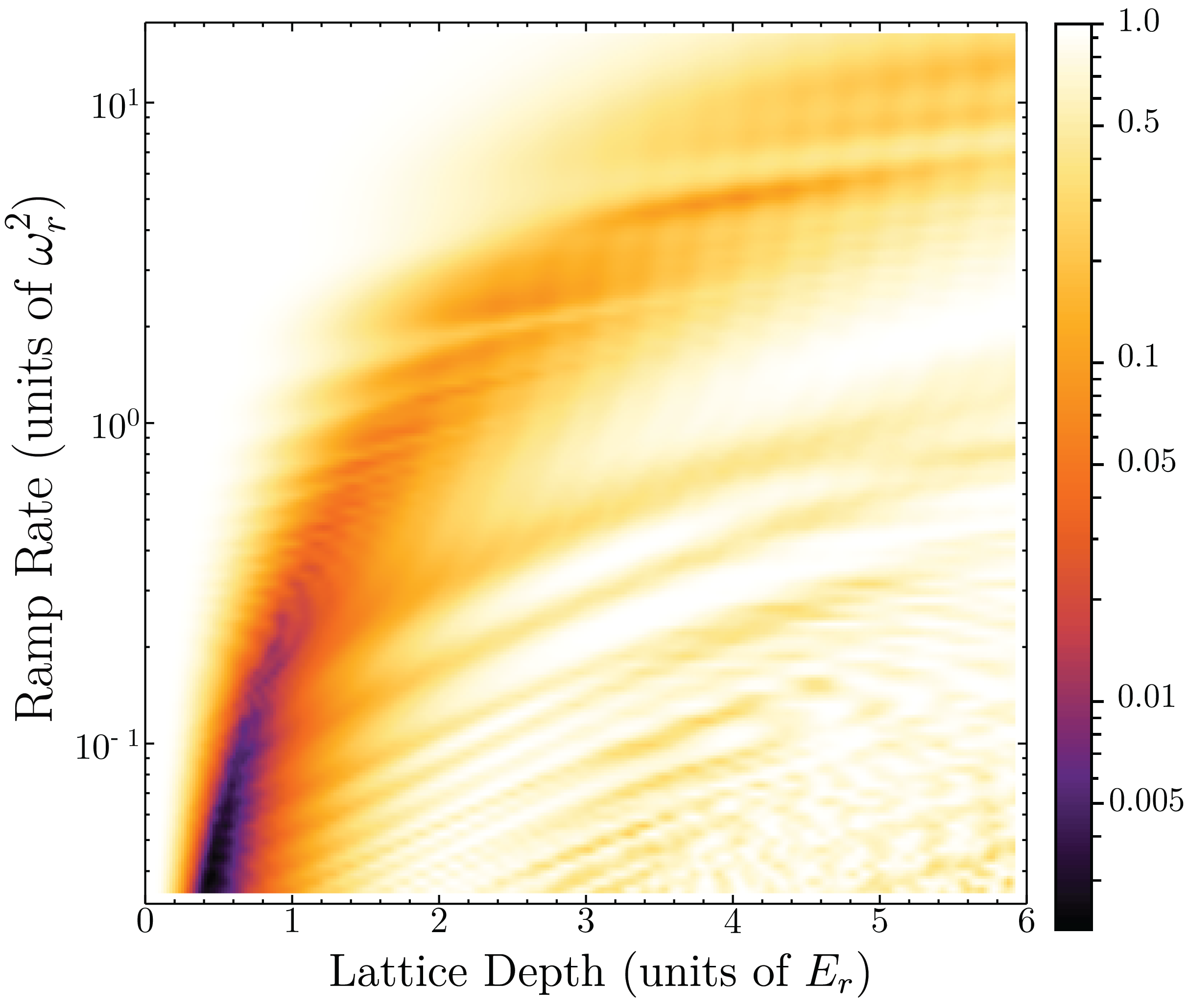}
\caption{Numerical simulation of beamsplitter losses showing the dependence on the two most important parameters in the Hamiltonian: frequency ramp rate and lattice depth. Darker color corresponds to lower losses, or higher efficiency. The simulation includes adiabatic loading of lattice, frequency ramping for four Bloch periods, and adiabatic unload, such that the final momentum splitting is $16 \hbar k$. Efficiency is defined as the probability amplitude on the desired momentum states after unloading the lattice. See text for discussion of loss mechanisms.}
\label{efficiency map}
\end{figure}

\subsection{\label{sec:Combined ramp}Comparison of limits on the ramp rate}

The RWA condition in Eq.~(\ref{reduced RWA condition}) and Landau--Zener tunneling losses both place an upper limit on the ramp rate. For Landau--Zener losses, efficient dynamics require $r\ll (\pi/2) U_0^2\hbar^2$; when $U_0\lesssim \sqrt{2} E_r$, the RWA condition in Eq.~(\ref{RWA2}) is automatically satisfied if the lattice depth is large enough to sufficiently suppress Landau--Zener tunneling. The RWA that leads to the Hamiltonian (\ref{final Ham}) is therefore asymptotically correct in the limit $r\rightarrow 0$ provided that $\hbar \sqrt{r}\ll U_0 \lesssim \sqrt{2} E_r$. On the other hand, when $U_0\gtrsim \sqrt{2}E_r$, both the RWA condition and the standard Landau--Zener criterion begin to fail because the time windows for successive transitions begin to overlap non-negligibly.

Higher-order losses place a lower limit on the ramp rate, and for $r\leq \omega_r^2$, this limit and the upper limits on the ramp rate from Landau--Zener losses and the RWA condition can all easily be satisfied. Because of the non-linear scaling of these different limits on the ramp rate, the maximum possible efficiency of the processes quickly approaches $1$ as $r\rightarrow 0$; for $r=0.5\omega_r^2$, the maximum efficiency of the initial $4\hbar k$ momentum splitting in a Bloch beamsplitter is already $>99\%$.

Fig. \ref{efficiency map} illustrates beamsplitter losses as a function of the ramp rate $r$ and the lattice depth $U_0$. Losses towards the top-left of the plot correspond to Landau--Zener tunneling losses, and losses towards the bottom-right correspond to higher-order transitions. Moving towards higher lattice depths and ramp rates, the maximum efficiency of the beamsplitter decreases because of the competing loss mechanisms.

The two loss channels result in non-zero wavefunction amplitude in momentum states different from the target states, and these additional momentum states could contribute to parasistic interferometers. This analysis is beyond the scope of the paper, however we note that there exist methods to reduce the effects of parasitic interferometers \cite{kasevich parasitic}.  




\subsection{\label{sec:degeneracy}Crossing through velocity degeneracy}

In addition to a beamsplitter, one can also ramp the two lattices through velocity degeneracy to create atom mirrors and combiners. This process has previously been attempted experimentally \cite{Sven degeneracy}, but the dynamics were seen to be inefficient and uncontrolled because the ramp rate, lattice depth, and relative phase between lattices were not optimized. The intuition for the dynamics through a level crossing are described below, and for a more mathematical treatment see Appendix \ref{sec:zero crossing}.

Consider two optical lattices with velocities that are initially far apart. One arm of an interferometer that is initially comoving with one of the two lattices can be understood as a superposition of an even-parity and an odd-parity ground state. Relative phase shifts between the even and odd states causes amplitudes to add constructively or destructively for positive or negative momentum states, which means that a controlled relative phase shift between the even- and odd-parity states can be used to control the momentum distribution of the atomic state after crossing through velocity degeneracy.

Figure \ref{fig:bands}c shows the band structure as the lattices are ramped through velocity degeneracy at time $t=0$. Far from velocity degeneracy, the even and odd ground state energy bands overlap and have the same level crossing structure. Near time $t=0$, however, 
these energy bands deviate because, by definition, an odd-parity state in momentum space cannot have amplitude on the zero-momentum basis state $\ket{0}$. As a result, when crossing through velocity degeneracy the odd-parity ground state has no level crossing coupling momentum into or out of the zero momentum state, so the even parity ground state passes through two additional level crossings at times $t = \pm T_B/2$ compared to the odd parity ground state.

Through the coherent interactions with photons from each of the lattices, the relative phase $\phi_0$ of the two optical lattices is ultimately added to amplitude in the even-parity state, but not the odd-parity state. As a result, the offset phase $\phi_0$ can coherently control the population in the two lattices after a degeneracy crossing. This allows one to create reflection or recombination pulses in an interferometer, and together with the beamsplitter process described previously, this comprises a full set of atom-optics tools for atom interferometry (see Fig.~\ref{fig:Reflection Transmission} for experimental implementation).

\subsection{\label{eh}Experimental considerations}

The dynamics of symmetric Bloch oscillations are sensitive to the initial velocity distribution of an atom. Efficient beamsplitter dynamics are observed for atoms with velocity spreads of more than $\sigma_v = 0.5 v_r$, where $\sigma_v$ is the standard deviation in velocity of a Heisenberg-limited Gaussian wavepacket. However, this spatial separation does not necessarily result in a superposition state in momentum space. For matter wave sources where different velocity classes are uncorrelated, only amplitude within a certain momentum window $\Delta p$ results in a superposition state, and amplitude to the left (right) of this window in momentum space will preferentially follow the right-moving (left-moving) lattice \cite{Berman BBS}. Intuitively, this can be understood by considering the dynamics in the Brillouin zone. When an atom begins at zero velocity, symmetric Bloch oscillations apply a force in both directions, and the quasimomentum can be thought of as being ramped in both directions simultaneously such that the state reaches both edges of the Brillouin zone at the same time, splitting the atom symmetrically in a superposition state. If the atom has some initial velocity, however, it will reach one edge of the Brillouin zone before the other, and as a result amplitude will preferentially be driven by this first transition. 

Numerical integration of the Hamiltonian (\ref{BBS Hamiltonian}) can be used to solve for evolution of a wavefunction $\psi(x,t)$ with arbitrary initial conditions (see Fig.~\ref{fig:fig1}) using the Crank--Nicolson method to discretize the Schr\"odinger Equation \cite{real space, real space new}. These simulations confirm that faster ramp rates result in higher fidelity superposition states in momentum states, which in turn results in higher contrast interferometers. 


Diffraction phases are fundamental to any asymmetric Bragg diffraction beamsplitter \cite{Diffraction phase, Brian diffraction phase}, and must be accounted for in precision measurements \cite{alpha}. For symmetric Bloch oscillations, if the center of the initial atomic velocity distribution is non-zero, the initial state has some projection onto the odd-parity eigenstates which leads to asymmmetry and diffraction phases. The symmetry of the Bloch beamsplitter (see Fig.~\ref{fig:fig1}) ensures that there is no diffraction phase that is fundamental to the technique. An initial velocity of the atoms, however, breaks the symmetry and creates a diffraction phase between interferometer arms. The numerical study discussed in Appendix \ref{sec:diffraction phase} shows that there are ``magic" lattice depths where the diffraction phase vanishes. For realistic experimental control over the stability of the lattice depth, the diffraction phase can be limited to $\pm 10\, \text{mRad}$, independent of the momentum splitting. Increasing the momentum splitting will therefore fractionally suppress the diffraction phase, and diffraction phases can also be measured directly by varying the time between pulses in an interferometer. Note also that an ensemble of atoms with different center velocities will result in phase spreading in an interferometer.

The analytic results derived for Landau--Zener tunneling and the rotating wave approximation only apply to slow ramp rates that satisfy the condition in Eq.~(\ref{reduced RWA condition}). Experimentally, we use larger ramp rates of up to $r = \,10\omega_r^2$ and lattice depths around $8E_r$ in order to maximize interferometer contrast, which is a region of parameter space that breaks the assumptions used to derive this inequality. Although the analytical efficiency predictions break down in this regime, we still observe reasonably efficient dynamics both numerically and experimentally. In fact, the velocity bandwidth of the beamsplitter is larger at faster ramp rates which results in higher contrast interferometers. See Fig. \ref{efficiency map} for an illustration of a beamsplitter for different values of lattice depth and ramp rate. Notably, even in regions of parameter space outside where the RWA is valid, one can still achieve relatively low loss beampslitters.

\section{\label{sec:experiment}Experiment}

Our experimental apparatus has been described previously in \cite{alpha}. A magneto-optical trap of Cesium atoms is launched vertically in an atomic fountain. The cloud is further cooled to a few hundred nK using polarization gradient cooling and Raman sideband cooling. Three successive Raman transitions prepare the atom in the internal state $\ket{F=3,m_F=0}$ with a vertical velocity spread around 0.05 recoil velocities $v_r$. 

The frequencies $\omega_1$ and $\omega_2$ in the Hamiltonian (\ref{BBS Hamiltonian}) are ramped in the lab to compensate for Doppler shifts from gravitational acceleration of the atoms such that in the atom's inertial frame, $\omega_1 = \omega_2 = \omega$. The optical lattice is detuned by +80 GHz (blue) from the Cs D2 line, and is formed from a roughly Gaussian beam with $1/e$ waist of about 3 mm that is retroreflected. The frequency components $\omega_1$ and $\omega_{2\pm}$ are cross-polarized and a quarter waveplate is placed in front of the retroreflecting mirror such that the desired lattices are formed upon retroreflection. The laser intensity is actively stabilized by feeding back to the drive power of an acousto-optic-modulator (AOM) \cite{Brian thesis}. 

\begin{figure}
\includegraphics[width=3.35in]{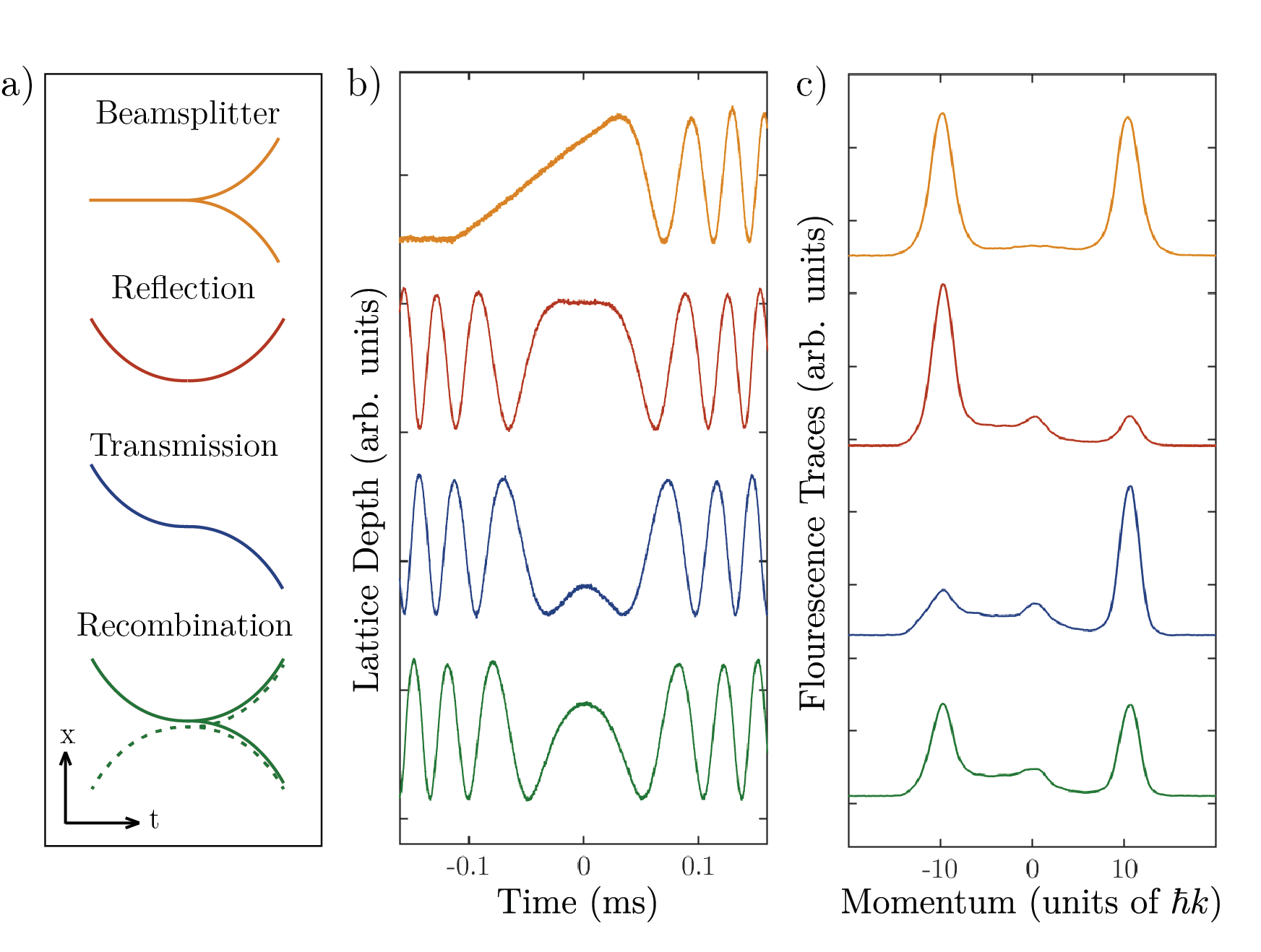}
\caption{Experimental realization of a Bloch beamsplitter (yellow), reflection (red), transmission (blue), and recombination (green) as lattices are ramped through velocity degeneracy. a) Space-time trajectories. b) Intensity profiles of the $\omega_2 \pm \omega_m(t)$ interferometry beams, which are measured by imaging the laser beams on a photodiode just before entering the vacuum chamber. The profiles show beats between the two frequencies, which is the temporal part of the potential in the Hamiltonian (\ref{BBS Hamiltonian}). Time $t=0$ indicates when $\omega_m\,=\,0$. Different phase offsets $\phi_0$ result in different beat profiles on the beam. c) Fluorescence traces of atoms from time-of-flight imaging showing the resulting distribution after various operations.}
\label{fig:Reflection Transmission}
\end{figure}

\begin{figure*}
\centering
\includegraphics[width=6.4in]{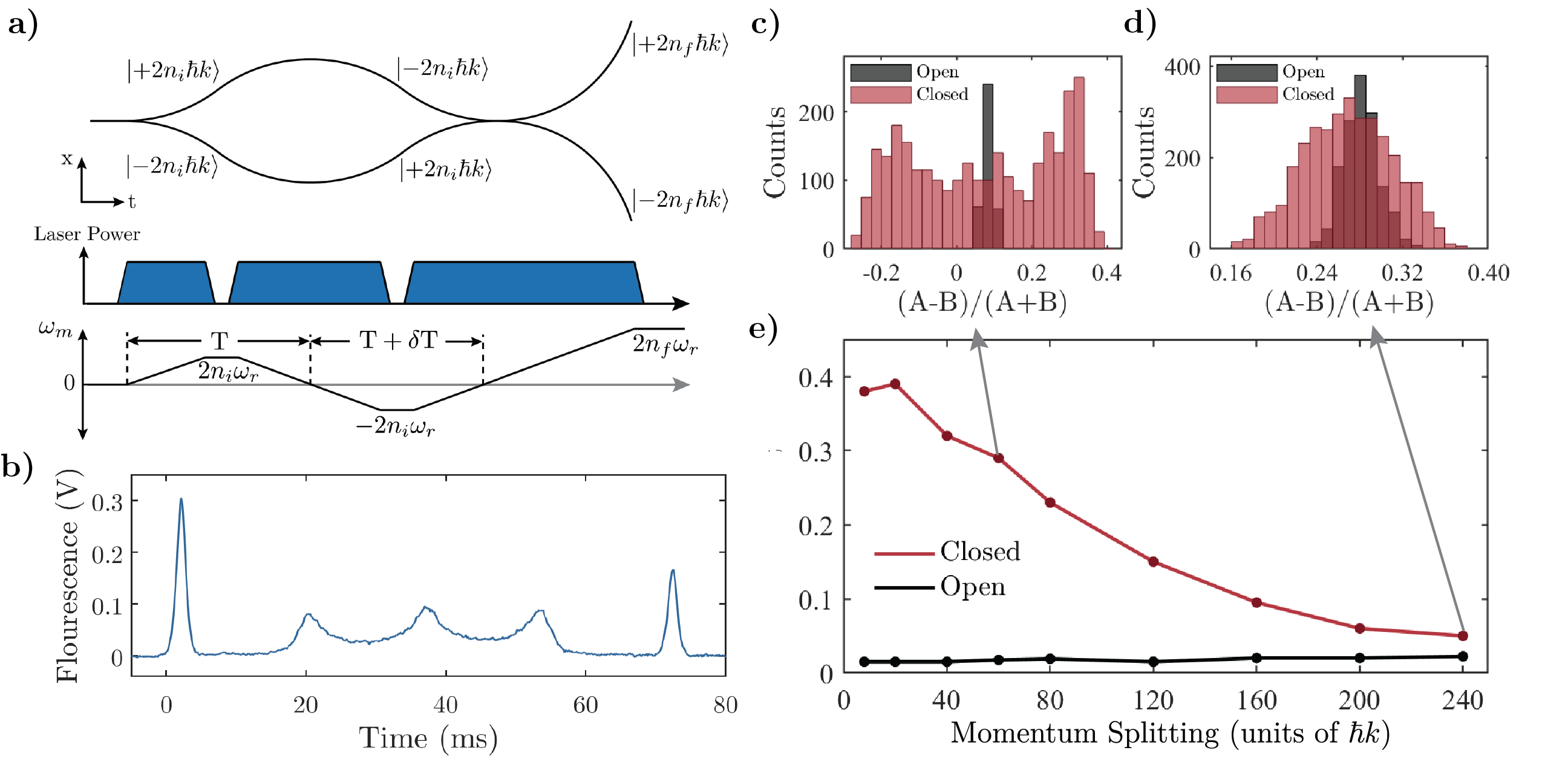}
\caption{Experimental realization of a Mach--Zehnder interferometer. a) Interferometer geometry, laser intensity profile, and profile of the modulation frequency $\omega_m$ vs. time. The time interval $T$ is defined based on $\omega_m$-zero crossings. A time offset can be used to open the interferometer to eliminate interference, while $\delta T = 0$ leads to maximum contrast. b) Sample fluorescence trace of a $T = 8.5$\,ms, 60$\hbar k$ MZ interferometer. c,d) Histogram of population fractions for 60$\hbar k$ and 240$\hbar k$ momentum splittings, respectively, in $T=8.5$\,ms interferometers. Population fraction is defined as $(A-B)/(A+B)$ where $A$ and $B$ are populations in the two output ports. e) Contrast versus momentum splitting for closed and open interferometers. For all data points, $\omega_m$ is ramped at a rate of $r/(2\pi)=249$ MHz/s.}
\label{MZ}
\end{figure*}

The modulation frequency $\omega_m(t)$ from the Hamiltonian in Eq.~\ref{BBS Hamiltonian} determines the velocity splitting between the two lattices. It is generated experimentally by mixing the output of an AD9959 digital frequency synthesizer with a 10-MHz clock and low-pass filtering the output, after which $\omega_m(t)$ is mixed into the drive frequency for an AOM to generate frequency sidebands that are written onto the laser. The offset phase $\phi_0$ is a tuneable parameter on the digital frequency synthesizer. 

\subsection{\label{sec:atom optics}Atom optics with symmetric Bloch oscillations}

To create a beamsplitter, atoms are adiabatically loaded into two velocity-degenerate lattices that initially add constructively to form a single lattice, which corresponds to $\omega_{m} = 0$ and $\phi_0 = 0$ at time $t=0$ in Eq.~(\ref{BBS Hamiltonian}). The modulation frequency is then ramped linearly at a rate $r$ such that $\omega_{m} = rt$ and the two lattices accelerate away from one another. The resulting momentum distribution is then measured using time-of-flight detection, as shown in Fig.~\ref{fig:fig1}b. For a $\pm2n$-photon beamsplitter, the final atomic state after the beamsplitter is mostly in the $\ket{\pm n}$ states, with a small number of atoms left in the $\ket{0}$ state. 

In addition to an initial beamsplitter, a full interferometer sequence requires reflection pulses to reverse momentum of the interferometer arms and a recombination pulse to interfere the two arms together. 
We find that varying the offset phase $\phi_0$ between the two lattices from $0$ to $\pi$ controls the population in the two lattices after the degeneracy crossing, varying from reflection to transmission with a beamsplitter/recombination behavior at an intermediate $\phi_0$. This phase also dictates the interference (``beat") between the two optical lattices at the time of the modulation frequency zero crossing, as shown in Fig.~\ref{fig:Reflection Transmission}b. The optimal phase offsets $\phi_0$ are found which maximize population in the desired output channels, as shown in Fig.~\ref{fig:Reflection Transmission}. Note that the optimal phase offset $\phi_0$ is dependent on both the ramp rate and the lattice depth $U_0$; the beat profiles shown in Fig.~\ref{fig:Reflection Transmission}b are specific to the lattice depth and ramp rate used experimentally, and will need to be optimized anew if either parameter is changed. This is due to the fact that the dynamical phase $\phi_d$ in Eq.~(\ref{final state}) is a function of both the ramp rate and the lattice depth. For the parameters used in our experiment, the simulated efficiencies are similar to those realized experimentally, but experimentally we see more atoms lost to the zero momentum state.

\begin{figure}[t]
\centering
\includegraphics[width=3.3in]{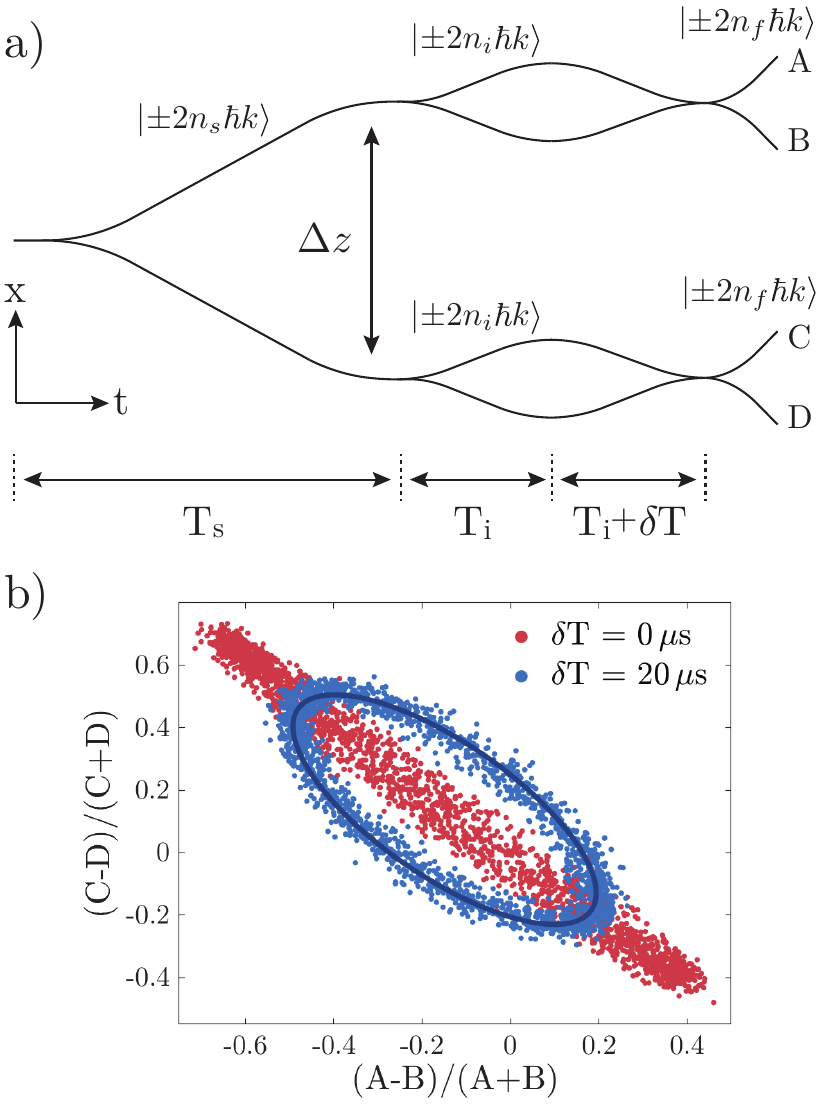}
\caption{a) Schematic of a dual-lattice gradiometer. $A$ and $B$ are populations in output ports of one MZ, and $C$ and $D$ are populations in output ports of the second MZ. b) Parametric plot of data taken using $n_s$ = 125 and $T_s$ = 150 ms for $\Delta z\: \approx \:$ 11 cm vertical splitting between the two MZ interferometers. Within the interferometers, $n_i\,=\,5$ and $T\,=\,10\,$ms, and $n_f$ = 10 to resolve the outputs. The dark line is an ellipse fit to the $\delta$T = 20$\,\mu$s data.}
\label{gradiometer}
\end{figure}

\subsection{\label{sec:MZ}Mach--Zehnder interferometer}

Combining these techniques, we implement a MZ interferometer, see Fig.~\ref{MZ}. The sequence starts with a Bloch beamsplitter that is ramped to some final momentum splitting $n_i$. After this, a reflection sequence is performed and the phase $\phi_0$ in the Hamiltonian Eq.~(\ref{BBS Hamiltonian}) is arranged as shown in Fig.~\ref{fig:Reflection Transmission}b. The two halves of the wavefunction are then interfered using a recombination sequence and the outputs are separated to some final momentum state $n_f$.

To optimize the contrast of the detected interferences, we need to separate signal atoms from background atoms that arise from loading and unloading the lattices. Using $n_f>n_i$ separates ``signal" atoms from those backgrounds in time-of-flight imaging, see Fig.~\ref{MZ}b. 

A ramp rate for $\omega_m$ of $r=2\pi*249$ MHz/s $= 9.3\,\omega_r^2$ and a lattice depth of around 8 recoil energies (for each lattice individually) are used, as these parameters resulted in the largest interferometer contrast. The phases $\phi_0$ for the two degeneracy crossings were also optimized experimentally to maximize contrast. In between different interferometer operations, we switch the direction of the modulation frequency ramp by switching RF frequency sources for the modulation frequency $\omega_m(t)$, and we adiabatically unload the lattice during this time to avoid losses from the ground state. 

We observe up to 40\% contrast in a $T=8.5$ ms, $20 \hbar k$ interferometer where atoms are guided in the lattices during 16.7 ms of the 17 ms interferometer duration (Fig.~\ref{MZ}e). Because of vibrational noise in the experiment, it was not possible to observe a stable fringe, so contrast was determined by measuring the fluctuations in the output populations on a histogram. Without changing the laser intensity profile, momentum transfer is increased by changing the profile of $\omega_m$ as shown in Fig.~\ref{MZ}a, and contrast is observed up to $240\hbar k$ momentum splitting. 

\subsection{\label{sec:Gradiometer}Gradiometer}

Observing contrast in an interferometer does not show that the interferometer is phase-stable. In order to show phase-stability and first-order coherence \cite{Holger reply}, we also perform a differential measurement between two MZ interferometers in a gradiometer configuration, see Fig.~\ref{gradiometer}. In this configuration, phase noise from vibrations is common to both MZ interferometers, so the differential measurement can reveal a stable relative phase. The two MZ interferometers are separated vertically by roughly 11 cm by using a 500$\hbar$k Bloch beamsplitter with a ramp rate of $r=2\pi*249$ MHz/s $= 1.9\,\omega_r^2$. Within each MZ, a momentum splitting of 20$\hbar$k, an interferometer time $T_i\,=\,10\,$ms, and a ramp rate of $r=2\pi*249$ MHz/s $= 9.3\,\omega_r^2$ are used. The slower ramp rate for the first beamsplitter minimizes background atoms in the time-of-flight traces, and the faster ramp rate during the interferometer maximizes contrast. Phase-stability is observed between the interferometers by plotting the relative populations parametrically (see Fig.~\ref{gradiometer}b). 

If there is no differential phase acquired between the interferometers, one sees perfect correlation in the outputs, and common-mode vibration noise causes data to fall at different points on this line. Instead, we see that the outputs are anti-correlated, owing to a $\pm \pi/2$ phase shift imprinted on the upper and lower MZ interferometers, respectively, during the opening pulse of the interferometer. Similar phases are well known in higher-order Bragg transitions, and come directly from Schrodinger equation dynamics \cite{Brian thesis}. Note that this phase is not permitted from symmetry arguments: the opening of the upper and lower interferometers around time $T_s$ are asymmetric, since upper (lower) arm has positive (negative) velocity prior to the splitting. 

Differential phase shifts between the two MZs results in an elliptical distribution in the parametric plot. We find that a timing delay $\delta T$ of the final recombination, as defined in Fig.~\ref{gradiometer}a), introduces a controlled phase difference into the interferometer that scales linearly with the timing delay, $\Delta \phi = (40$\,rad/ms$) \delta t$. Differences in gravity between the two MZs also creates a differential phase shift which is proportional to the gravity gradient. However, this phase is around 5 mrad for the parameters used experimentally and is too small to be observed. The phase coherence between the two MZs demonstrates that the technique is first-order coherent and phase-stable, and can therefore be used for measurements in atom interferometry. We achieve as large as 50\% contrast in the differential measurement, which is similar to the largest contrast we ever observed with Bragg diffraction in the same instrument. The contrast is higher than the contrast in the Mach--Zehnder interferometers in Fig.~\ref{MZ} because the lattice in the gradiometer configuration is turned off when the lattices are not being accelerated. The timing delay causes loss of contrast because of not fully closing the interferometer.


\section{\label{sec:conclusion}Conclusions and Outlook}

We have developed new techniques for coherently controlling superpositions of momentum states by generalizing Bloch oscillations to two independently accelerated optical lattices. First, the Hamiltonian was treated analytically, and it was shown that the dynamics can produce efficient and coherent atom optics elements, even when the lattices pass through velocity degeneracy. For slow ramp rates, the process is adiabatic and atoms can adiabatically follow the even-parity ground state of Hamiltonian (\ref{final Ham}). When ramping lattices through velocity degeneracy, the populations in the two lattices can be controlled by changing the relative phase of the two optical lattices, allowing for all atom-optics elements required to form an interferometer. Using only accelerated lattices, we create LMT interferometers with high contrast, and we showed that the resulting dynamics were first-order coherent using a differential measurement.


Compared to existing atom optics techniques \cite{102hk,BBB,New Abend LMT}, DLBO offer a number of advantages. Applications with constraints on laser power and free-fall distance, such as space-based interferometry \cite{NASA CAL, BECCAL} or portable gravimeters \cite{minig}, can use these techniques to maximize momentum transfer and thus sensitivity. 
Being based on adiabatic processes, these methods are robust to fluctuations in experimental parameters like lattice depth or laser frequency \cite{Clade}. 
Symmetric Bloch oscillations are more robust to small laser intensity variations than Bragg diffraction beam splitters, and can eliminate systematic phase shifts known as diffraction phases \cite{Estey 2015, Parker PRA, Gupta PRA}. Moreover, large momentum transfer can be obtained with modest laser power, whereas in multi-photon Bragg diffraction the required laser intensity scales proportional to $n^2$ or even $n^4$, if scattering losses are to be kept constant \cite{holger Bragg theory}. In contrast, the laser power required for DLBO is independent of the momentum splitting, relaxing the laser power requirements in an experiment. 
Compared to combinations of Bragg diffraction and Bloch oscillations \cite{BBB,Abend Thesis}, DLBO requires less laser power and can achieve higher efficiencies. For example, two sequential $4\hbar k$ double-Bragg beamsplitters used in reference \cite{New Abend LMT} use a peak lattice depth of $3-4 E_r$ and achieve a total efficiency around 90\%, and higher-order double Bragg pulses require considerably more laser power. In contrast, the $60\hbar k$ beamsplitter in Fig.~\ref{fig:fig1}b uses a lattice depth of $1.5 E_r$ while achieving an efficiency greater than 90\%. 

A generalization of these dual-lattice techniques shows promise for new measurements of the fine-structure constant $\alpha$. A set of realistic experimental parameters are outlined in Appendix \ref{sec:alpha}, where we show that $10^8$ radians of phase are attainable. This paves the way for a measurement of alpha at the $10^{-11}$ level, an order of magnitude improvement on existing measurements. Another generalization of the Bloch beamsplitter uses a multi-photon, $4n\hbar k$ transition to open the interferometer where $n>1$. Our numerical simulations show that this multi-photon process also leads to an efficient beamsplitter for appropriate ramp rates and lattice depths, see Appendix \ref{sec:higher order} for further discussion. 


The authors thank Matt Jaffe, Victoria Xu, Sven Abend, Ernst Rasel, Justin Khoury, and Tanner Trickle for useful discussions and feedback, as well as the past group members who helped develop the experimental apparatus including Brian Estey, Joyce Kwan, Chenghui Yu, Pei-Chen Kuan, and Shau-Yu Lan. This work was supported by the National Science Foundation Grant No. 1806583, the National Institute of Science and Technology Grant No. 60NANB17D311, and the W.M. Keck Foundation Grant No. 042982. Z.P. acknowledges funding from the National Science Foundation GRFP. 

\section{\label{sec:appendix}Appendices}

\subsection{\label{sec:SLBO Ham}Unitary transformation for single-lattice Bloch Hamiltonian}

In an inertial frame initially comoving with the atoms, the SLBO Hamiltonian can be written as:

\begin{multline}
    \label{Ham Bloch lab frame}
    H = \sum_{l=-\infty}^{\infty}\Bigg(\frac{(2 l \hbar k)^2}{2 m}\ket{l}\!\!\bra{l} \\ + U_0 e^{i\left(\frac{r t^2}{2} + \phi_0\right)}\left(\ket{l}\!\!\bra{l+1} + \ket{l}\!\!\bra{l-1}\right)\Bigg)
\end{multline}
The Hamiltonian, Eq.~(\ref{Ham Bloch}), is derived by transforming this Hamiltonian, Eq.~(\ref{Ham Bloch lab frame}), into a rotating frame that puts the time dependence of the rotating terms into the diagonal. This is achieved with the following unitary:
\begin{align}
    \label{unitary Bloch}
    U &= \sum_{l=-\infty}^{\infty} e^{i \frac{d(t) \hat{p}}{\hbar}}e^{i \frac{\theta(t)}{\hbar}}\ket{l}\!\!\bra{l} \\ &= \sum_{l=-\infty}^{\infty} e^{i l \left( \frac{r t^2}{2} + \phi_0\right)}e^{i\frac{m a^2 t^3}{6 \hbar}}\ket{l}\!\!\bra{l}
\end{align}
with $d(t) \equiv a t^2/2 + \phi_0/k$ and $\theta(t) \equiv m a^2 t^3/6$. This same transformation is used in reference \cite{Clade 2}, and it is almost identical to the transformation used in Eq.~(\ref{unitary}), except there is no longer a absolute value sign on the momentum operator. Acting on the Hamiltonian, Eq.~(\ref{Ham Bloch lab frame}), with the unitary transformation in Eq.~(\ref{unitary Bloch}) results in $H_{\text{SLBO}}$:

\begin{multline}
    \label{Ham Bloch}
    H_{\text{SLBO}} = \sum_{l = -\infty}^{\infty}\frac{(2 l \hbar k - F t)^2}{2 m}\ket{l}\!\!\bra{l} \\ + \frac{U_0}{2}\left(\ket{l}\!\!\bra{l+1} + \ket{l}\!\!\bra{l-1}\right)
\end{multline}
The $F t$ term that appears in the kinetic energy is related to the quasimomentum $k_q$ through the relation  $\hbar k_q = F t$.

\subsection{\label{sec:symmetrized}Symmetrized Hamiltonian}

The Hamiltonian in Eq.~(\ref{final Ham}) can be explicitly symmetrized by applying a rotation to the basis states. This is achieved by rotating to new basis states that are symmetric and antisymmetric combinations of the free-space momentum basis states, namely we will have (unnormalized) even parity basis states $\ket{+_l} = \ket{l} + \ket{-l}$ and odd parity states $\ket{-_l} = \ket{l} - \ket{-l}$. The zero momentum state remains unchanged under this rotation, as it is already an even-parity state. The following rotation matrix achieves this transformation:

\begin{multline}
    \label{rotation}
    R = \ket{0}\!\!\bra{0} + \sum_{l>0} \frac{1}{\sqrt{2}}\left(\ket{l}\!\!\bra{l} + \ket{-l}\!\!\bra{-l}\right) \\ + \frac{1}{\sqrt{2}}\left(\ket{l}\!\!\bra{-l} - \ket{-l}\!\!\bra{l}\right)
\end{multline}

The Hamiltonian (\ref{final Ham}) can then be rotated to the symmetric Hamiltonian $H_{\text{sym}} = R\, H_{\text{DLBO}}\, R^T$ to arrive at the following:

\begin{multline}
    \label{rotated Ham}
    H_{\text{sym}} = \frac{(F t)^2}{2 m}\ket{0}\!\!\bra{0} \\ + \sum_{l>1}\bigg(\frac{(2 |l| \hbar k - F t)^2}{2 m}\left(\ket{+_l}\!\!\bra{+_l} + \ket{-_l}\!\!\bra{-_l}\right) \\ + \frac{U_0}{2}\Big(\ket{+_l}\!\big(\bra{+_{l+1}} + \bra{+_{l-1}}\big) + \ket{-_l}\!\big(\bra{-_{l+1}} + \bra{-_{l-1}}\big)\!\Big)\!\!\bigg)\\ + \frac{U_0}{2}\left(\ket{+_1}\!\!\bra{+_2} + \ket{-_1}\!\!\bra{-_2}\right) + \frac{U_0}{\sqrt{2}}\left(\ket{0}\!\!\bra{+_1} + \ket{+_1}\!\!\bra{0}\right)
\end{multline}
In this rotated basis, there is no coupling between $\ket{0}$ and $\ket{-_1}$, so we can explicitly see why the odd-parity states have no level crossing at times $t=\pm T_B/2$ in Fig.~\ref{fig:bands}c). Moreover, the coupling between $\ket{0}$ and $\ket{+_1}$ is $\sqrt{2}$ larger than any of the other couplings, resulting in suppressed Landau--Zener tunneling from the level-crossings of the even-parity ground state at times $t=\pm T_B/2$ in Fig.~\ref{fig:bands}c).

\subsection{\label{sec:RWA}Rotating wave approximation condition}

To make the rotating wave approximation (RWA) in Eq.~(\ref{new Ham}), we average the oscillating term $e^{i r t^2}$ over the duration of the transition between momentum states. This term is oscillating most slowly around the first level crossing between the first and second even bands at time $t=T_B/2$. In the limit of small lattice depths $U_0\ll 4 E_r$, the energy gap $E_g(t)$ near this level crossing is given by

\begin{equation}
    \label{energy gap}
    E_g(t) = \sqrt{\hbar^2 r^2(t - T_B/2)^2 + 2 U_0^2} \;\;\; ,
\end{equation}
such that the center of the level crossing occurs at time $t=T_B/2$, and the duration of the level crossing is $\Delta t = 2\sqrt{2}U_0/\hbar r$.

Taking the time average of the rotating term $e^{i r t^2}$ over the duration of the level crossing gives the following:

\begin{equation}
    \label{RWA1}
    \langle e^{irt^2}\rangle 
    \approx -\frac{i \hbar^2 r}{4U_0}e^{i\alpha}\frac{U_0\cos\beta-2\sqrt{2}i E_r\sin\beta}{8 E_r^2-U_0^2}
\end{equation}
where we define $\alpha=2(8 E_r^2+U_0^2)/(\hbar^2 r)$ and $\beta=8\sqrt{2} E_r U_0/ (\hbar^2 r)$, and we have assumed that $r\ll 2(2\sqrt{2} E_r-U_0)^2/\hbar^2$. The rotating term can be dropped so long as this average is small compared to 1, i.e., when 
\begin{equation}
    \label{RWA2}
    |\langle e^{irt^2}\rangle|<\frac{\hbar^2 r}{4U_0(2\sqrt{2} E_r-U_0)}\ll 1
\end{equation}
or equivalently, $r\ll 4U_0(2\sqrt{2} E_r - U_0)/\hbar^2$. We note that varying the time window of integration in Eq.~(\ref{RWA1}) changes the numerical factors in Eq.~(\ref{RWA2}), but not the limiting behavior as $r\rightarrow 0$.

\subsection{\label{sec:higher order}Higher-order loss mechanisms}

\begin{figure}[t]
\centering
\includegraphics[width=3.3in]{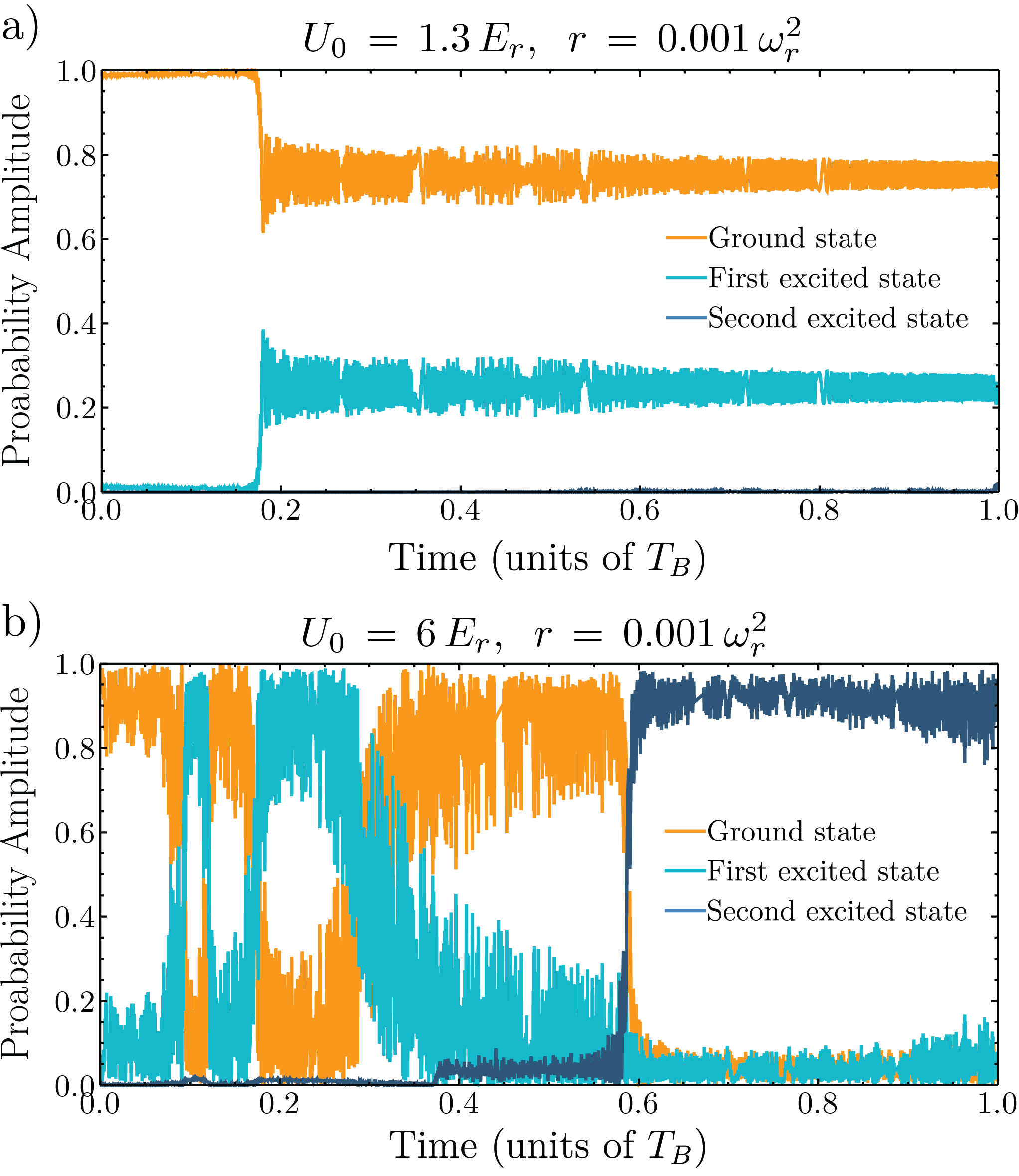}
\caption{Simulations of one Bloch period of a Bloch beamsplitter illustrating losses from the ground band due to higher-order transitions. The states used for determining the probability amplitude are even-parity eigenstates of Hamiltonian (\ref{final Ham}). A slow ramp rate is used so that the various transitions are resolved from one another. a) The first losses to occur are due to a third-order transition coupling the ground state and first excited state. b) A much larger lattice depth shows a number of different higher-order transitions. Before time $t=T_B/2$ there are four separate higher-order resonances between the ground state and first excited state that transfer population between the levels. Around time $t=0.6\,T_B$ there is a transition between the ground state and the second excited state.}
\label{Higher order}
\end{figure}

When the lattice depth is too large, the oscillating terms dropped in the rotating wave approximation from the Hamiltonian in Eq.~(\ref{new Ham}) can contribute to higher-order parasitic transitions. The dominant loss mechanism at ramp rates $r\ll \omega_r^2$ is a third-order (six-photon) process coupling the states $\ket{0}$ and $\ket{+_1}$ around time $t=T_B/6$, where $\ket{+_l}$ refers to the symmetrized basis states derived in Appendix \ref{sec:symmetrized}. There are two possible energy and momentum conserving pathways for the transition to occur; $\ket{0} \rightarrow \ket{+_1} \rightarrow \ket{0}\rightarrow\ket{+_1}$ and $\ket{0}\rightarrow \ket{+_1}\rightarrow\ket{+_2}\rightarrow\ket{+_1}$. For lattice depths much less than the spacing between energy levels, $U_0/2 \ll 4 E_r$, the effective coupling between these states scales like $(U_0/2)^3/(4E_r)^2$, which is the same scaling as the Rabi frequency in higher-order Bragg diffraction \cite{holger Bragg theory, Chenghui Thesis}.

During a Bloch beamsplitter, the laser frequencies are swept across this parasitic resonance, as seen in Fig.~\ref{Higher order}a, which can be thought of as a parasitic level-crossing between $\ket{0}$ and $\ket{+_1}$; for an efficient Bloch beamsplitter, amplitude should remain in $\ket{0}$ by tunneling through this level-crossing diabatically. To first order, the adiabatic population transfer to the state $\ket{+_1}$ during this level crossing is given by $P_{LZ} = 1 - e^{-2\pi\Gamma} \approx 2\pi\Gamma$ when the Landau--Zener parameter $\Gamma$ is close to zero. For $U_0 \ll 8 E_r$ and $r\ll\omega_r^2$, we therefore expect losses from the Bloch beamsplitter $P_{\text{loss}} = 2\pi\Gamma_3 \propto \left(\omega_r^2/r\right)\left(U_0/8E_r\right)^6$ where $\Gamma_3 \propto \left(\omega_r^2/r\right)\left(U_0/8 E_r\right)^6$. This scaling of the higher-order losses in the limit of $r\rightarrow 0$ agrees with our numerical simulations.

In addition to the third-order process discussed above, there are an infinite number of these higher-order processes that conserve energy and momentum, but the transition rates are highly suppressed at lower lattice depths. Fig.~\ref{Higher order}b show the result of a simulation with an increased lattice depth, to a regime in which many of these higher-order transitions can couple amplitude to higher-excited states. The parameters chosen for this simulation happen to drive five of these higher-order transitions within the first Bloch period. A ramp rate $r\ll\omega_r^2$ is chosen for the simulation so that the transitions are well-resolved. In contrast, Fig.~\ref{fig:LZ}a) illustrates negligible higher-order losses because all higher-order transitions are highly suppressed at lower lattice depths. 

\subsection{\label{sec:zero crossing}Crossing through velocity degeneracy}

The dynamics while crossing through velocity degeneracy are determined by studying the eigenstates of the DLBO Hamiltonian, Eq.~(\ref{final Ham}). An initial momentum state $\ket{n}$, where $n>0$, can be decomposed as
\begin{equation}
    \label{initial state}
    \ket{n} = \frac{1}{\sqrt{2}}(\ket{+_n} + \ket{-_n})
\end{equation}
where 
are the symmetric and antisymmetric combinations of the free-space momentum basis states $\ket{\pm n}$ as derived in Appendix \ref{sec:symmetrized}. Similarly, $\ket{-n}$ can be decomposed as
\begin{equation}
    \label{initial state minus}
    \ket{-n} = \frac{1}{\sqrt{2}}(\ket{+_n} - \ket{-_n})
\end{equation}
Without loss of generality, we restrict our attention to one arm of an interferometer with momentum $\ket{n}$. Then when one of the two lattices is initially comoving with the state $\ket{n}$, this state will be loaded into the ground state of the DLBO Hamiltonian in Eq.~(\ref{final Ham}) as a superposition of odd-parity and even-parity ground states according to Eq.~(\ref{initial state}). 

Crucially, relative phase shifts between the even- and odd-parity eigenstates cause amplitude to add constructively or destructively for the positive momentum or negative momentum states; for example, if the state $\ket{-_n}$ acquires a $\pi$ phase shift relative to the state $\ket{+_n}$, then the state $\ket{n}$ in Eq.~(\ref{initial state}) will transform the the state $\ket{-_n}$ in Eq.~(\ref{initial state minus}). There are two sources of relative phase shifts between the even- and odd-parity states as the lattices are swept through velocity degeneracy. First, since these states are energy eigenstates of the Hamiltonian, there is a dynamical phase difference $\phi_d$ between the two states given by $\phi_d = (1/\hbar)\int dt' (E_{-}(t') - E_{+}(t'))$, where $E_{\pm}$ denotes the energy of the even- and odd-parity ground states over time, as shown in Fig.~\ref{fig:bands}c. Since the even- and odd- parity states have different level structure near the degeneracy crossing, this gives a non-trivial phase shift. In addition, there are two additional level crossings for the even state near velocity degeneracy compared to the odd-parity state, as discussed in Sec.~\ref{sec:symmetrized}. These level crossings correspond to transferring photons to and from the laser field, so the phase of the laser field is imparted to the atomic state during these crossings. 

Laser phase is a well known source of phase in atom interferometers, and is the primary phase contribution for certain interferometer configurations such as Mach--Zehnder interferometers \cite{Chu laser phase}. In a single optical lattice, laser phase arises when the position of the laser standing wave shifts position with respect to the atom, resulting in a phase shift $\Delta \phi = 2 k \Delta x$. In the case of two optical lattices, there is an additional degree of freedom, namely the relative position of the two lattices. This changes the offset phase $\phi_0$ in Eq.~(\ref{BBS Hamiltonian}), and it is reasonable to expect this phase term to play a coherent role in the dynamics.

There are two ways to understand the laser phase effects, mathematically and physically. Mathematically, one can see that the even-parity state is shifted relative to the odd-parity state from the definition of the unitary transformation in Eq.~(\ref{unitary}). As mentioned previously, the sign on $d(t)$ in Eq.~(\ref{unitary}) is changed at time $t=0$, which changes the phase offset on every basis state except for the zero momentum state $\ket{0}$. Just before time $t=0$, the odd-parity ground state is approximately given by $\ket{-_{gs}} = (e^{-i \phi_0}\ket{1} - e^{-i \phi_0}\ket{-1})/\sqrt{2} = e^{- i \phi_0}\ket{-_1}$, whereas after time $t=0$ the state becomes $\ket{-_{gs}} = (e^{i \phi_0}\ket{1} - e^{i \phi_0}\ket{-1})/\sqrt{2}  = e^{i \phi_0}\ket{-_1}$. The odd state is therefore phase shifted by $2\phi_0$. Since the state $\ket{0}$ is unchanged, the even- and odd-parity states see a relative phase shift of $2\phi_0$. Physically, the nature of the degeneracy crossing is a result of constructive or destructive interference between amplitudes. Since there are two additional level crossings of the even state compared to the odd state, the even state receives a laser phase shift $\phi_l = 2\phi_0$. At time $t=T_B$, after the two additional crossings, both the even- and odd-parity states are mostly superpositions of the states $\ket{\pm l}$, but the extra phase shift of the even state results in coherent interference and changes the resulting output state. This phase shift can also be observed in our numerical simulations, where the even-parity state is phase shifted by $\phi_0$ at each of the two level crossings near velocity degeneracy. 


Up to a global phase, the new state after the degeneracy crossing can be written as: 
\begin{equation}
    \label{final state}
    \ket{\psi_f} = \frac{1}{\sqrt{2}}(e^{i (\phi_d + \phi_l)}\ket{+_n} + \ket{-_n})
\end{equation}
By controlling the phase shifts $\phi_d$ and $\phi_l$ in an experiment, one has control over the output nature of the degeneracy crossing. For example, arranging for $\phi_d + \phi_l = 2m\pi$ for some integer $m$ ensures that the state after the crossing will be identical to the state before the crossing, which corresponds to transmission through the crossing. For $\phi_d + \phi_l = (2m+1)\pi$ for some integer $m$, the output state becomes $-\ket{+_n} + \ket{-_n} = \ket{-n}$, which has opposite momentum compared to the input state $\ket{n}$ and corresponds to a reflection. Intermediate values of the phase can be used to split amplitude between the two momentum states $\ket{\pm n}$. In practice, it is easiest to change $\phi_0$, and therefore $\phi_l$, since this phase is directly controllable experimentally. Our simulations show that $\phi_d$ also depends on $\phi_0$ at the moment that the lattices are velocity degenerate, but this dependence does not prevent one from continuously transforming between different output behaviours by changing only $\phi_0$.

The phase $\phi_d$ is dependent on the lattice depth, and therefore the lattice depth needs to be well controlled in order to see coherent dynamics after the zero-crossing. In the limit $U_0 = 0$, the dynamical phase $\phi_d$ is given by $\phi_d = 16 \omega_r^2/r$, such that $\phi_d \gg 2\pi$ when $r\ll\omega_r^2$. When $U_0 >0$, this phase term is also a function of the lattice depth; as a result, fluctuations in $U_0$ lead to fluctuations in $\phi_d$. Similarly, variable $U_0$ across a finite laser beam leads to a variable $\phi_d$ across an atom cloud. Both of these effects result in unreliable zero-crossing behaviour at slow ramp rates, and both effects likely explain why we see the largest interferometer contrast for fast ramp rates around $r=10\omega_r^2$.

\subsection{\label{sec:diffraction phase}Diffraction phase}

\begin{figure}[t]
\centering
\includegraphics[width=3.3in]{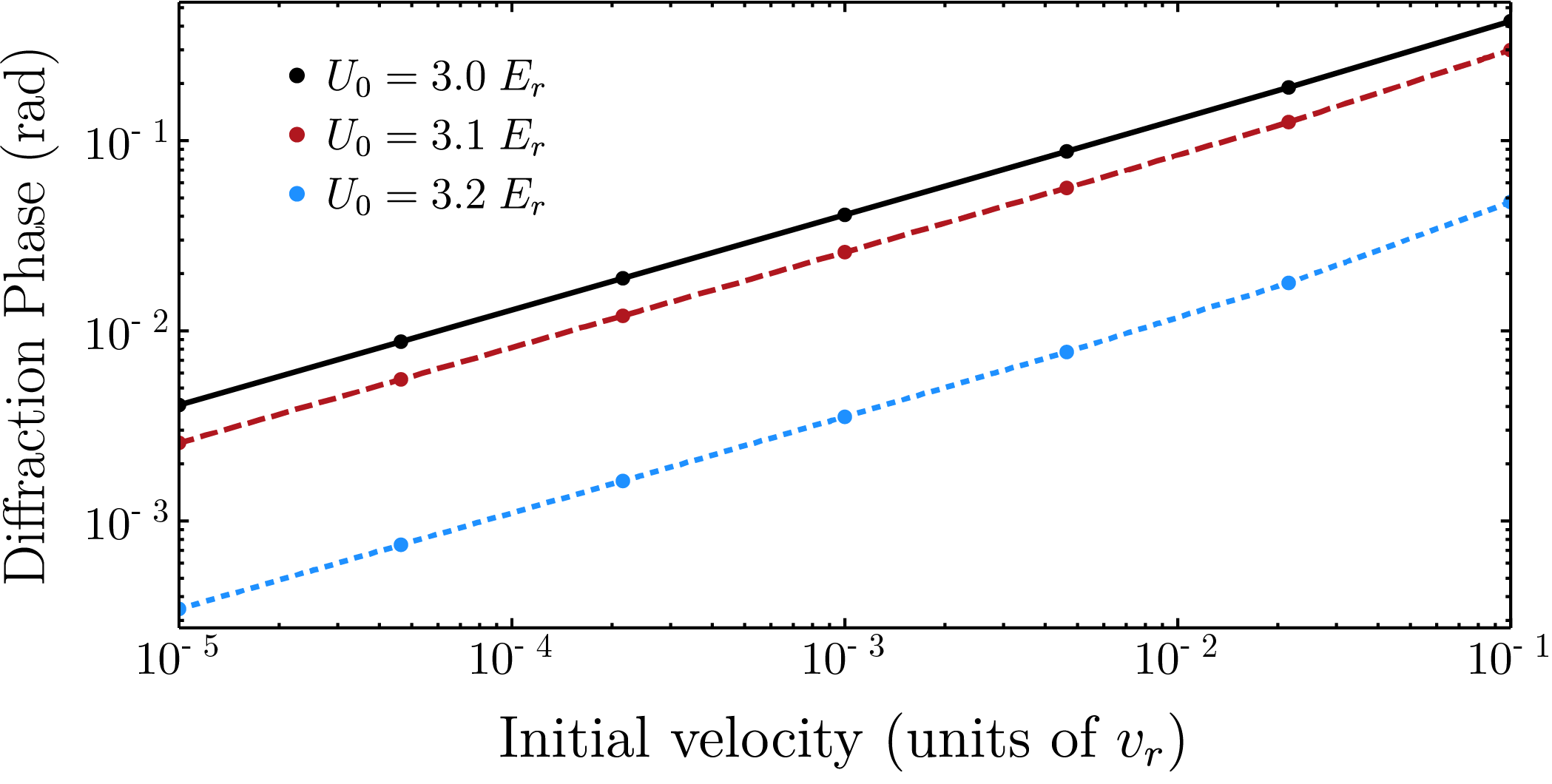}
\caption{Numerical simulation of diffraction phase from a Bloch beamsplitter as a function of velocity with respect to the initial optical lattice. Simulation includes adiabatically loading the lattice, frequency ramping at a rate $r=1.0\, \omega_r^2$ for four Bloch periods, then adiabatic unloading of the lattice. See text for further discussion.}
\label{diffraction phase}
\end{figure}

Here, we consider the diffraction phase acquired from a beamsplitter, which is the phase difference between the positive and negative momentum components of the resulting wavefunction. If the atomic state initially has some free-space velocity with respect to the lattice, the momentum-parity symmetry of the problem is broken and the resulting dynamics will be asymmetric, leading to a diffraction phase. 

\begin{figure}[t]
\centering
\includegraphics[width=3.3in]{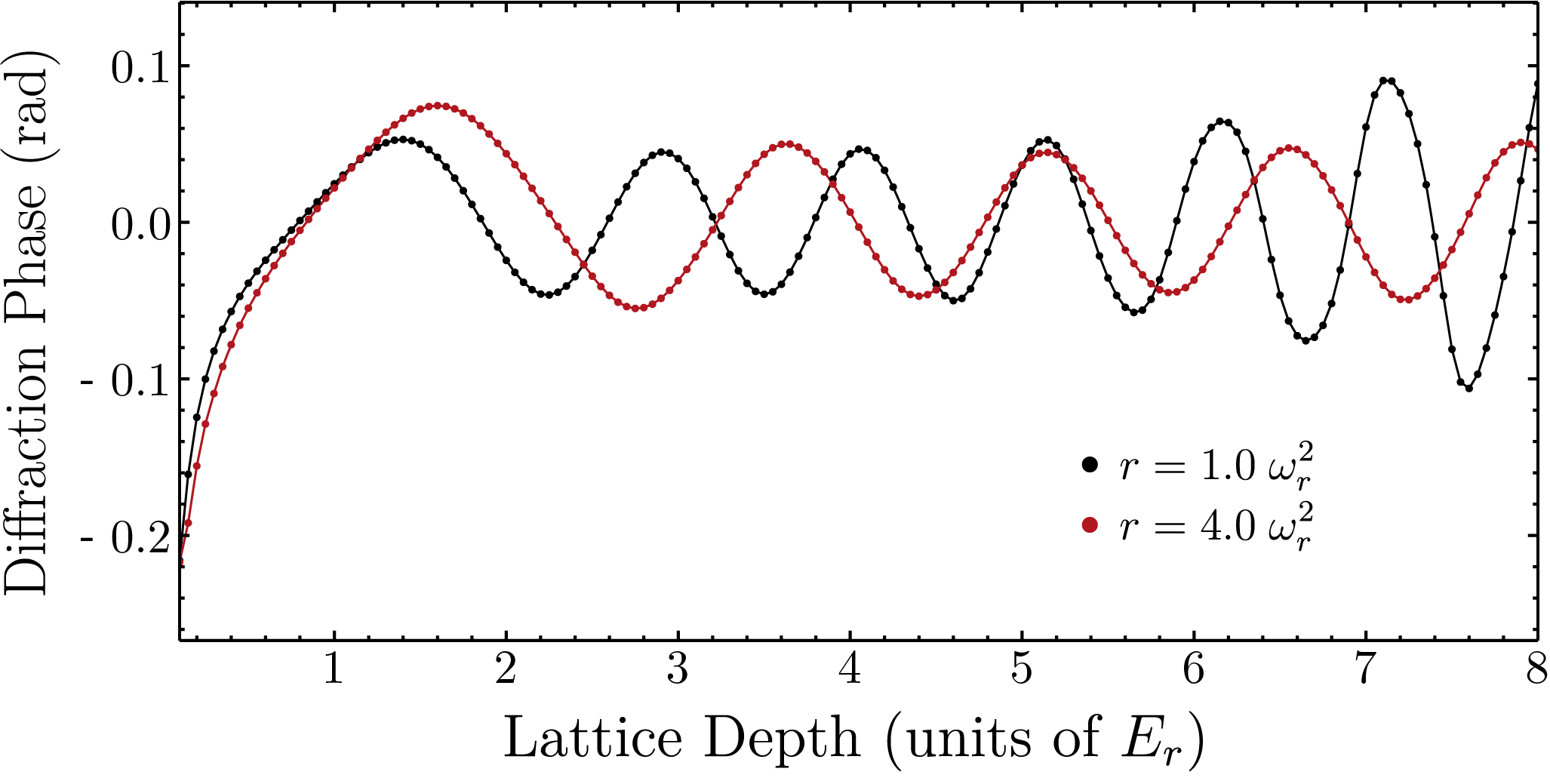}
\caption{Numerical simulation of diffraction phase from a Bloch beamsplitter as a function of lattice depth for two different ramp rates, using an initial velocity with respect to the lattice of $0.001 v_r$. Points are from simulation, lines are an interpolation between points to guide the eye. The zero crossings in the diffraction phase allow for one to operate an interferometer at a ``magic" lattice depth to suppress sensitivity to diffraction phase. See text for further discussion.}
\label{diffraction phase 2}
\end{figure}

Figures \ref{diffraction phase} and \ref{diffraction phase 2} show numerical simulations of the diffraction phase for a $16\hbar k$ Bloch beamsplitter. 
Almost all of the diffraction phase from the beamsplitter comes from the first $8\hbar k$ momentum splitting near velocity degeneracy; further increasing the momentum transfer beyond this does not further increase the dynamical phase $\phi_d$. The diffraction phase for a beamsplitter scales like the square root of the initial velocity, but the prefactor in front of this scaling can be controlled by varying the lattice depth and the details of loading or unloading the lattice. The simulations in Figures \ref{diffraction phase} and \ref{diffraction phase 2} use a linear intensity ramp for loading an unloading over a time $t_{\text{load}} = 6\pi \omega_r^{-1}$.

Figure \ref{diffraction phase 2} shows the diffraction phase as a function of the lattice depth, and oscillations in the diffraction phase allow one to operate at a ``magic" lattice depth with suppressed sensitivity to diffraction phases from missing the center velocity of the atom cloud. For precision measurement, such magic lattice depths could be used to significantly reduce the diffraction phases caused by fluctuations in experimental parameters. For example, a ramp rate of $r=4\omega_r^2$ and a lattice depth around $U_0 = 5.9 E_r$ gives 80\% efficient beamsplitters with minimized diffraction phase sensitivity (see Figures \ref{efficiency map} and \ref{diffraction phase 2}). We can reasonably operate within $0.001 v_r$ of the center velocity of the atom cloud, and by intensity stabilizing the lattice to 1\% fluctuations, the diffraction phase can be limited to $\pm 10$ mRad. This diffraction phase can then be measured directly by varying the duration of the interferometer, as done in reference \cite{alpha}.

\subsection{\label{sec:higher order beamsplitter}Higher-order generalization of the dual-lattice methods}

The transitions driven in DLBO are two-photon processes that transfer $2\hbar k$ momentum. By sweeping past multiple of these transitions in successions, LMT can be easily achieved. In contrast, higher-order transitions are also possible that transfer $2n\hbar k$ momentum in a single, multi-photon process. 

It is instructive to first understand single-lattice higher-order processes before understanding the dual-lattice analogues.  SLBO can be though of as adiabatically sweeping past a successions of $2\hbar k$ Bragg transitions \cite{Unitary}. The higher-order, multi-photon analogue has been implemented experimentally in reference \cite{Kovachy Bragg ARP}. The laser is adiabatically swept across a $2 n\hbar k$ Bragg resonance, which adiabatically drives a $2n$-photon process. Though not discussed directly in \cite{Kovachy Bragg ARP}, this process can be interpreted using a Bloch band picture where atoms have an initial quasimomentum outside of the first Brillouin zone such that they are loaded into higher Bloch bands. As the lattice is accelerated, the state sweeps past a level crossing between higher Bloch bands, and successful momentum transfer requires the state to adiabatically traverse the crossing and stay in the same Bloch band.

DLBO can be thought of as adiabatically sweeping past a succession of ``double Bragg" transitions \cite{double Bragg}. A first-order double Bragg transition symmetrically drives $\pm 2\hbar k$ Bragg resonances such that the two arms are split by $4\hbar k$ momentum. One can also symmetrically drive two higher-order Bragg resonances that transfer $\pm 2 n \hbar k$ momentum to obtain a $4n\hbar k$ beamsplitter, as are implemented in references \cite{Abend Thesis, New Abend LMT}.

It is also possible to adiabatically sweep past a higher-order double Bragg transition. In terms of the modulation frequency $\omega_m$ in Eq.~(\ref{BBS Hamiltonian}), these resonances occur at $\omega_m = (2 m+1)\omega_r$ for integers $m$. A $4n\hbar k$ adiabatic dual-lattice beamsplitter can be achieved by sweeping past one of these resonances adiabatically. An experimental sequence would consist of the following: 1) atoms are adiabatically loaded into a lattice with a modulation frequency slightly below the desired resonance, 2) the modulation frequency is swept across the resonance, and 3) the atoms are adiabatically unloaded from the lattice. It is important that the modulation frequency does not become close to other resonances during this sequence. Unlike a Bloch beamsplitter, continued ramping of $\omega_m$ after a high-order beamsplitter process will not transfer more momentum, but rather alternate between increasing and decreasing the momentum splitting between arms. The average momentum transfer per Bloch period will still be $4\hbar k$, as in the ground band. 

Our simulations of this process show that it can be more efficient than a Bloch beamsplitter at a given ramp rate.  
However, there are two major downsides to these higher-order dual-lattice techniques. First, much more laser power is required to drive the transition; the power required to drive an $n^{\text{th}}$-order Bragg transition scales sharply with the order $n$, namely as $n^2$ to maintain the same Rabi frequency, and $n^4$ to also maintain the same single-photon scattering rate \cite{holger Bragg theory}. Second, continued ramping of the lattices does not continue to increase momentum splitting in any advantageous way compared to using the ground band. As a result, the first-order dual-lattice methods discussed in the main text are easier to use if the goal is to achieve very large momentum splitting without the need for significantly more laser power.

\subsection{\label{sec:alpha}Application to recoil measurements}

\begin{figure}[t]
\centering
\includegraphics[width=3.3in]{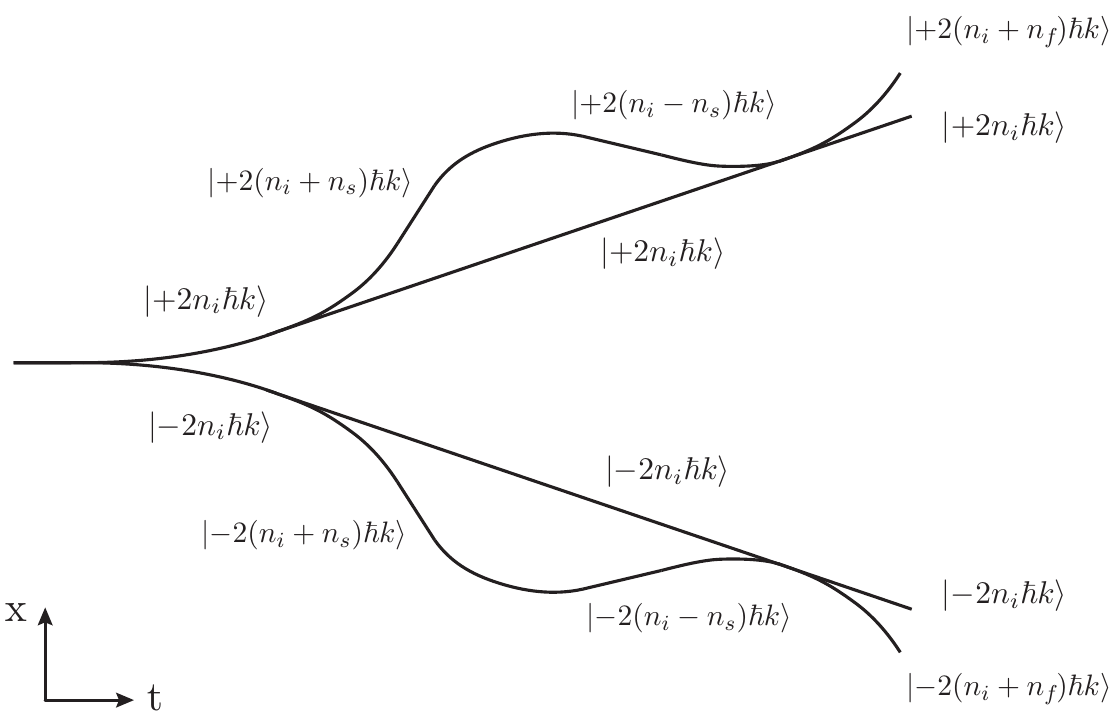}
\caption{Interferometer geometry sensitive to an atomic recoil phase. The asymmetry between the two upper (lower) trajectories leads to a kinetic recoil phase acquired by the upper (lower) interferometer. The simultaneous conjugate interferometer configuration is used for a differential measurement that cancels gravitational phase to fist order, and adds the recoil phases in the upper and lower interferometers. Addressing the four velocity classes of light requires one left-moving frequency and four right-moving frequencies, similar to Fig. \ref{fig:levels}a).}
\label{recoil fig}
\end{figure}

A generalization of DLBO shows promise for atom recoil measurements, and therefore in measurements of the fine-structure constant $\alpha$ \cite{alpha}. This section is included as an example of the potential applictaions of DLBO, however we note that before such a measurement, many new systematic effects would likely need to be studied. 

By removing the assumption that $\omega_1$ = $\omega_2$ and are independent of time in the Hamiltonian in Eq.~(\ref{BBS Hamiltonian}), asymmetric lattice guided geometries can be created \cite{Kovachy guided interferometer}. Additional light frequencies can also be added to the laser in order to address more than two velocity classes of atoms at the same time. Figure \ref{recoil fig} shows an example interferometer configuration that would be sensitive to an atom recoil phase. The phase in the interferometer can be calculated by integrating the energy of the atoms over the various trajectories \cite{Chenghui Thesis}. Assuming that the time to accelerate atoms is much less than the time between beamsplitter or reflection pulses, the phase of the interferometer is given by 

\begin{equation}
    \phi = 16 \omega_r n_s^2 T
\end{equation}

where $\omega_r$ is the recoil frequency of the matter wave, $n_s$ is defined in Fig. \ref{recoil fig}, and T is the time between beamsplitter and reflection pulse in the upper (or lower) interferometer.

The following outlines a set of realistic experimental parameters that could lead to $10^8$ radians of recoil phase, an order of magnitude improvement in sensitivity over the leading recoil measurement \cite{alpha}.  Based on the results discussed in Section \ref{sec:experiment}, atoms in our apparatus can interact with up to $1000$ photons inside an interferometer where contrast can still be observed. Choosing $n_i = 100$, $n_s = 80$ and $n_f = 100$ (defined in Fig.~\ref{recoil fig}) requires atoms to interact with $840$ photons before closing the interferometers. For the calculation, we use a time of $80$ ms between opening the interferometers and slowing the arms back to having the same velocity, the same timing used in reference \cite{alpha}. Using a frequency ramp rate of $r=250$ MHz/s, Cesium atoms can be accelerated from $\ket{2 n_i \hbar k}$ to $\ket{2 (n_s + n_i) \hbar k}$ in roughly 6 ms, which is much less than the time between different pulses. This ramp rate was shown to give good interferometer contrast in the main text for atoms with a vertical velocity spread of $0.05 v_r$.

\end{document}